\documentclass[a4paper,10pt]{article}
\usepackage{jheppub} 

\usepackage{physics}
\RequirePackage{verbatim}
\NewDocumentCommand{\tens}{t_}
{%
	\IfBooleanTF{#1}
	{\tensop}
	{\otimes}%
}
\NewDocumentCommand{\tensop}{m}
{%
	\mathbin{\mathop{\otimes}\displaylimits_{#1}}%
}
\usepackage{amsmath}
\usepackage{color}
\usepackage{amsfonts}
\usepackage{verbatim}
\usepackage[toc,page]{appendix}
\usepackage{amssymb}
\usepackage{bbold}
\usepackage{graphicx}
\usepackage{pgfplots}
\usepackage{float}

\usepackage{amsfonts,amssymb,amssymb,physics,,hyperref}
\usepackage{xfrac}
\usepackage{tensor}
\usepackage{mathtools}
\usepackage{dsfont}
\usepackage[english]{babel}
\usepackage{amsthm}

\usepackage[all]{xy} %to draw commutative diagrams of mapping

\usepackage[thinc]{esdiff}
\usepackage{mathtools}
\usepackage{geometry}
\usepackage{filecontents}
\usepackage[hang,flushmargin]{footmisc}
\usepackage{cancel}
\usepackage{scalerel}
\usepackage{fancyhdr}

\definecolor{brightgreen}{rgb}{0.0, 0.5, 0.0}
\hypersetup{
	colorlinks=true,
	linkcolor=blue,
	filecolor=magenta,      
	urlcolor=blue,
	citecolor=brightgreen
} % To accepts extra parameters to set up the final PDF file like blue hyperref(links)

\makeatletter
\def\overUnderArrow{\@ifnextchar[\overUnderArrow@i{\overUnderArrow@i[]}}
\def\overUnderArrow@i[#1]#2#3{% #1 under #2 over #3 main argument
	\ifx\relax#1\relax\array[b]{c}\overset{\text{#2}}{\uparrow}\\#3\endarray
	\else\ifx\relax#2\relax
	\array[t]{c}#3\\\underset{\text{#1}}{\downarrow}\endarray
	\else
	\array{c}\overset{\text{#2}}{\uparrow}\\#3\\\underset{\text{#1}}{\downarrow}\endarray
	\fi\fi}
\makeatother
\newcommand{\cg}{\textnormal{\textsl{g}}}
\makeatletter
\renewcommand\@makefntext[1]{%
	\noindent\makebox[10pt][r]{\@makefnmark}#1}
\makeatother

\newcommand\numberthis{\addtocounter{equation}{1}
	\tag{\theequation}}
\newcommand{\myquad}[1][1]{\hspace*{#1em}\ignorespaces}
\newcommand{\del}{\partial}

\newcommand{\deldel}[3]{\frac{\del}{\del {#1}_{#2}^{#3}}}

\newcommand{\db}[2]
{\left[#1,#2\right]_{\scaleto{\text{D.B.}\mathstrut}{4.6pt}}}

\newcommand{\cmnt}[1]{\ignorespaces}
\renewcommand{\comm}[2]{\left[ #1,#2 \right]}

\DeclareUnicodeCharacter{03BA}{$\kappa$ }
\title{\boldmath Fate of $\kappa$-Minkowski space-time
	in non-relativistic (Galilean) and ultra-relativistic (Carrollian) regimes}
\author[1,a,b]{Deeponjit Bose,}\note{Corresponding author}
\affiliation[a]{Department of Physics, Ramakrishna Mission Vivekananda Educational and Research Institute\\PO-Belur Math, Howrah-711202, West Bengal, India\footnote{Bulk of this work was completed here}}
\affiliation[b]{Institute of Theoretical Physics, Jagiellonian University, ul. Prof. S. Łojasiewicza 11, 30-348 Krak\'{o}w, Poland}
\author[c,d]{Anwesha Chakraborty,}
\affiliation[c]{School of Mathematics and Statistics, The University of Melbourne, Parkville, VIC 3010, Australia}
\affiliation[d]{S.N. Bose National Centre for Basic Sciences\\
	JD Block, Sector III, Salt Lake, Kolkata-700106, West Bengal, India}
\author[a]{Biswajit Chakraborty}
\emailAdd{deeponjit.bose@doctoral.uj.edu.pl}
\emailAdd{anwesha.chakraborty@unimelb.edu.au}
\emailAdd{dhrubashillong@gmail.com}

\abstract{We present an algebraic and kinematical analysis of non-commutative $\kappa$-Minkowski spaces within Galilean (non-relativistic) and Carrollian (ultra-relativistic) regimes. Utilizing the theory of Wigner-In\"{o}nu contractions, we begin with a brief review of how one can apply these contractions to the well-known Poincar\'{e} algebra, yielding the corresponding Galilean and Carrollian algebras as $c \to \infty$ and $c\to 0$, respectively.
	Subsequently, we methodically apply these contractions to non-commutative $\kappa$-deformed spaces, revealing compelling insights into the interplay among the non-commutative parameters $a^\mu$ (with $|a^\nu|$ being of the order of Planck length scale) and the speed of light $c$ as it approaches both infinity and zero. Our exploration predicts a sort of ``branching" of the non-commutative parameters $a^\mu$, leading to the emergence of a novel length scale and time scale in either limit.
	Furthermore, our investigation extends to the examination of curved momentum spaces and their geodesic distances in appropriate subspaces of the $\kappa$-deformed Newtonian and Carrollian space-times. We finally delve into the study of their deformed dispersion relations, arising from these deformed geodesic distances, providing a comprehensive understanding of the nature of these space-times.}
\begin{document}
	\maketitle
	\flushbottom
	\numberwithin{equation}{section}
	\section{Introduction}
	Research in the realm of quantum gravity typically focuses on the ``ultra-relativistic regime," attempting to reconcile the interplay between strong gravity and quantum effects. In scenarios with a large number of uncorrelated degrees of freedom, quantum effects become negligible, and classical descriptions, as outlined by general relativity in the context of strong gravity, are expected to emerge while maintaining a finite speed of light. On the other hand, if the space-time happens to have some quantum nature, then some features of quantum gravity can also be seen in some macroscopic systems (see for example,   \cite{scholtz2008thermodynamics}). A notable example illustrating this kind of quantum space-times are found in non-commutative space-time models   \cite{DOPLICHER199439,JMadore_1992}, which formalize the anticipated fuzzy behavior of space-time near the Planck length ($L_p=\sqrt{\frac{\hbar G}{c^3}}\,\sim 10^{-35} $m). At this extremely small scale, quantum-gravitational effects challenge the conventional depiction of space-time as a smooth pseudo-Riemannian manifold.\\	
	Many of these non-commutative models establish a ``\emph{modified special relativistic regime}" within the framework of quantum gravity, where attempts are made to study the consequences of incorporating an additional scale(s) such as the Planck length $L_p$. Examples include theories like ``\emph{double or triple special relativity}"   \cite{amelino_doubly,AMELINO_CAMELIA_2002,amelino_dsr_its_results}. As the non-commutativity parameter, linked to the Planck length, approaches zero, the traditional description of Minkowski space-time as a commutative manifold is restored.\\\\	
	There is, however, an urgent need to investigate the persistence of residual quantum-gravity effects in the non-relativistic ($c\to\infty$) and ultra-relativistic ($c \to 0$) limits. Examining these limits is crucial due to their potential phenomenological implications. The $c\to \infty$ limit is relevant to the search for quantum-gravity signatures in systems where typical velocities are much smaller than the speed of light, as observed in scenarios like atomic interferometry   \cite{GARAY_1999,Ellis:1983jz}. Conversely, the $c\to 0$ limit with $L_p$ diverging ($L_p \,\to\,\infty$) can be formally associated with quantum effects in a strong gravity regime   \cite{dautcourt1998ultrarelativistic,Bergshoeff_2017}. Delving into these limits can provide valuable insights into the behavior of quantum gravity under extreme conditions and may contribute to our understanding of its phenomenological consequences.\\
	Galilean group capturing the non-relativistic picture of the real world dynamics is found in the limit $c\to\infty$ of the Poincare symmetry group and, in turn comes out to be the isometry of Newton-Cartan manifold $\mathcal{N}$   \cite{duval1985bargmann,duval_carroll,andringa2011newtonian,duval2009non} having a degenerate metric. Interestingly, the Galilean groups can be realized not only in the massive   \cite{weinberg_qft1}, but also massless cases   \cite{souriau2012structure,townsend_tachyons}. In fact the Bargmann algebra can be obtained as a central extension of the Galilean algebra with the mass $m$ playing the role of central charge   \cite{bagchi2009galilean}. The conformal extension of Galilean algebra (GCA) is found to be isomorphic to the BMS group   \cite{Bondi:1962px,sachs1962asymptotic,Barnich,alessio2018structure} (an infinite-dimensional symmetry group which is also the isometry of asymptotically flat space-time) in 2 and 3 dimensions   \cite{bagchi2010correspondence,bagchi2012bms} and has  its applications to condensed matter physics. Recently, a geometric description of the non-relativistic expansion of General Relativity (GR) in inverse powers of the speed of light $c$ has been developed, as documented in \cite{VandenBleeken:2017rij,Hansen:2019pkl,Hansen:2020pqs}, building on the prior work of \cite{Dautcourt:1996pm}. This advancement has been largely driven by new insights \cite{andringa2011newtonian,Christensen:2013lma,Hartong:2015zia,Hansen:2019vqf} into Newton–Cartan geometry, which supplants the Lorentzian geometry of GR at the leading order in the expansion. The resulting action for non-relativistic gravity, which is the next-to-next-to-leading order action in the expansion, incorporates Newtonian gravity but extends beyond it to account for strong gravitational fields, such as those causing gravitational time dilation. Overall, these expansion methods establish a foundation for a covariant and off-shell formulation at any order, making them particularly relevant to post-Newtonian expansions.
	\\ On the other hand, it is also intriguing to consider the opposite scenario where $c$ is very small. As first explored in \cite{levy_caroll}, the Poincar\'{e} group contracts to the Carroll group when the speed of light approaches zero, which in turn comes out to be the isometry group of a novel Carroll manifold $\mathcal{C}$   \cite{levy_caroll,ciambelli2019carroll,duval_carroll}, also having a degenerate metric and can be thought as a degenerate counterpart of the Galilei group as discussed in   \cite{levy_caroll,SenGupta:1966qer}. This results in some unusual consequences for kinematics and dynamics in this limit. Unlike the Galilean case, where light cones open up, the light cones close up in the Carroll limit. Consequently, particles with non-zero energy cannot move through space, and correlations between spatially separated events vanishes. Its study in GR, as initiated in \cite{Henneaux} and more recently \cite{Henneaux:2021yzg,hansen,Campoleoni}, offers new insights into its geometry and gravitational dynamics. More broadly, examining the small speed of light expansion of GR \cite{Dautcourt:1997hb} allows for a perturbative expansion around the (singular) Carroll point, analogous to the post-Newtonian expansion for large $c$.
	Carroll symmetry is also found by restricting a Lorentzian space-time to a null hyper-surface \cite{duval_carroll}. Carroll symmetries has found its application in the physics of a black hole horizon and in fact the horizon of black holes forms a Carroll manifold ($\mathcal{C}$)   \cite{marsot2022anyonic,donnay2019carrollian}. There have been a lot of studies on Carrollian sector in various fields like in gravitational waves and memory effects \cite{Duval:2017els,Zhang:2017rno}, in the Hall effect \cite{Marsot:2022imf}, in the hydrodynamics \cite{Ciambelli:2018wre,Petkou:2022bmz}, in quadratic and higher derivative gravity theories \cite{Tadros:2023teq,Tadros:2024qlo} etc.  Several conformal extensions of Carroll symmetry which are found to form subgroups of the BMS group \cite{Ciambelli:2019lap}, which in turn play important roles in gravitational physics \cite{duval2014conformal}. Further it has been shown in \cite{duval2014conformal} that the future null conformal boundary $\mathcal{I}^+$ of an asymptotically
	flat space-time emitting gravitational radiation is a Carroll manifold and its asymptotic symmetries, correspond to the elements of the BMS group.\\\\
	After establishing the importance of exploring non-relativistic and ultra-relativistic limits within the context of quantum gravity theories and considering their potential applications, we turn our attention to the specific case to be investigated in this paper. Our focus centers on the examination of $\kappa$-Minkowski space-time, chosen as a prototype of this framework for which we systematically analyze both its non-relativistic and ultra-relativistic limits, and subsequently derive the associated deformed dispersion relations.
	\\ \\
	$\kappa$-Minkowski space-time is one of the extensively studied toy model of  non-commutative space-time 
	that arises in the context of certain approaches to 
	quantum gravity \cite{Amelino-Camelia1,Daszkiewicz:2007ru,Young,Sitarz:2003kappa}, such as quantum deformations of space-time 
	symmetries. This model was proposed initially in   \cite{Luk,MAJID1994348,zak,luk1,sitarz} and later again in   \cite{amelino_doubly,amelino_dsr_its_results,glik,camelia2} in the context of double special relativity. This involves a parameter $\kappa$ having the dimensions of energy/ mass, and it sets 
	the scale for the non-commutativity of space-time 
	coordinates. Here, on the contrary, we consider a more general type of non-commutative space-times where the space-time coordinates 
	are taken to be operator-valued, fulfilling the following Lie algebra structure 
	:
	\begin{align}
		\label{def_minkow_eq1}
		\comm{\widehat{X}^\mu}{\widehat{X}^\nu} = i\theta^{\mu\nu}
		=i\left(a^\mu \widehat{X}^\nu - a^\nu \widehat{X}^\mu \right)
	\end{align}
	where components of $a^{\mu}$ are the non-commutative constant parameters. Note that despite its appearance $a^{\mu}$ is not taken here as a Lorentz four vector (as in   \cite{biswajit2023symmetries,wohlgenannt,wohlgenannt2,juric}), rather each of these 4 components are taken to be independent constants and, are Lorentz scalars where each of them are of the order of $\kappa^{-1}$ or inverse Planck mass (refer to   \cite{biswajit2023symmetries} for details). Nevertheless, we can still introduce their ``\emph{covariant}" counterparts formally as $a_\mu := \eta_{\mu\nu}a^{\nu}$ i.e. $a_0=a^0$
	and $a_i=-a^i \, (i=1,2,3)$.
	Not only that, we can go ahead to introduce a time-like, null and space-like $a^\mu$, just as for Lorentz 4-vectors.
	\\ 
	Clearly, in these kinds of spaces like in $\kappa$-Minkowski space-time the manifest  Lorentz covariance is lost, which can lend itself to modification of Lorentz transformations in the form of deformed action of the Poincare generators on the space-time coordinates and eventually impacting even the phase space(i.e. the Heisenberg) algebra:
	\begin{align*}
		\numberthis
		\label{eq12}
		\comm{\widehat{X}^\mu }{\widehat{P}_\nu}
		&=
		i\tensor{E^{-1}\left(\widehat{P};a\right)}{^\mu_\nu}
	\end{align*}
	from which we can retrieve back the usual Heisenberg algebra as $(E^{-1})^{\mu}\,_{\nu}\,\to\,\delta^{\mu}\,_{\nu}$ in the commutative limit : $a^{\mu}\,\to\,0$   \cite{biswajit2023symmetries,glickman_curved_mom_space}. Above equation \eqref{eq12} reveals a plausible and novel Planck scale effects, where coordinate translations become momentum-dependent-an important aspects of `relative locality'   \cite{smolin,glickman_curved_mom_space}. This can introduce a non-linear (non-commutative, non-associative) composition law for the momenta of particles and can be a possible signature for the associated curved momentum space. This aspect was studied later again in   \cite{Majid}, who coined, the term ``co-gravity" to refer to the curved momentum space. In fact, such a feature in the context of quantum gravity was foreseen way back in 1939 by Max Born   \cite{born}. Clearly here the space-time geometry becomes intimately inter-twined with the momentum space of particles, indicating an intricate relationship between energy-momentum and the fundamental structure of space-time. Further, as was indicated in   \cite{biswajit2023symmetries,glickman_curved_mom_space}, one can demote \eqref{eq12} along with the coordinate algebra  \eqref{def_minkow_eq1} to the level of classical Dirac brackets which denotes a deformed symplectic structure of the phase space arising from a certain first order form of the Lagrangian, describing the motion of a massive, spinless relativistic particle moving in $\kappa$-Minkowski space-time. Here, the deformation parameter is ``reduced" to $\mathfrak{a}^\mu$ with the associated dimension of inverse mass $\kappa^{-1}$ and is of the order of $\sqrt{\frac{\hbar}{G}}$ in the limit where both $\hbar, G\to 0$, holding their ratio fixed. This should apply to systems, having length scales $L\gg L_p=\sqrt{\frac{\hbar G	}{c^3}}$ and mass scale smaller than the Planck mass $m_p$. The right-hand side of the classical version of \eqref{eq12} then be interpreted as the momentum space vielbein/tetrads and one can use these tetrads to establish a classical map between the non-commutative coordinates $(X^{\mu})$ and commutative coordinates $(q^{\mu})$ as
	\begin{equation*}
		\numberthis
		\label{tetrad}
		X^{\mu}=\left(E^{-1}(P;a)\right)^{\mu}\,_{\alpha}\,\,q^{\alpha}
	\end{equation*}
	This is a \textit{non-canonical} map, the so-called Bopp map for $\kappa$-Minkowski space-time relating commutative to non-commutative coordinates. From \eqref{eq12}, the mapping can also be read-off using the undeformed Heisenberg algebra. At this juncture, the components of the momentum space tetrad assume significant importance for further analysis of momentum space geometry, given in terms of deformed metric and geodesic distance and eventually its impact on the deformed single particle dispersion relation and its potential impact on the laws governing addition of momenta in multi-particle system.\\\\
	In fact in a recent paper by some of us   \cite{biswajit2023symmetries}, the deformed Poincare symmetry of $\kappa$-Minkowski space-time was explored, revealing the emergence of non-trivial momentum space geometry and its impact on the deformed single particle dispersion relation. It was observed that the bare mass '$m$' gets renormalized to say $M$, and whose nature depends on the nature of $\mathfrak{a}^\mu$'s. Particularly, it was observed that for time-like and space-like $\mathfrak{a}^\mu$'s the corresponding power series expansions have a superficial resemblance to the power series expansions of relativistic dispersion relations in the ultra-relativistic(the so-called Carrollian ($c\to 0$)) and non-relativistic(the so-called Galilean($c\to\infty$)) limits respectively. This motivates us to delve deeper into this formal analogy and take this forward and to study the Galilean and Carrollian limits, of $\kappa$-Minkowski space-time. Our aim is to examine their impacts on the respective momentum space geometries on dispersion relations, if any. And, this is particularly motivated by the fact that in these two extreme limits $c\to\infty$ or $c\to 0$, both the space-time and the momentum space metrics become degenerate and one cannot have the corresponding limits of a time-like geodesics in momentum space for finite $c$ case as in   \cite{biswajit2023symmetries}. Furthermore, the momentum space metric, which is usually read-off from the quadratic Casimir operator, as in   \cite{biswajit2023symmetries} may not go over to another quadratic Casimir operator involving energy and spatial momentum components in these extremal cases. In fact, the quadratic forms will be shown to survive only in the spatial and temporal subspaces only, enabling us to read-off the metric components in the respective subspaces. It thus becomes quite imperative to obtain the corresponding deformed versions and to investigate the relationship between the deformed geodesic distance in appropriate subspaces and deformed dispersion relation from the first principle.
	\\ \\
	The paper is organized as follows: In section-2, as a part of a review, we have revisited the Poincare algebra and its non-relativistic ($c\to \infty$) and ultra-relativistic ($c\to 0$) contractions to Galilean and Carroll algebra respectively using space-time scaling method, where we have also derived the representation of the respective algebra generators. In section-3, we have introduced the notion of $\kappa$-Minkowski space-time algebra and used the above mentioned contractions to get its non-relativistic and ultra-relativistic counterparts of space-time algebra. Using the Bopp map of the $\kappa$-Minkowski space-time we make use of the contraction method in the phase-space variables to get the corresponding Bopp map for $\kappa$-Galilean and $\kappa$-Carrollian space-time in section-3.1 and 3.2 respectively.  We then used the maps to derive the deformed phase-space algebra for respective $\kappa$ deformed space-time and then used them in turn to obtain the vielbeins in the momentum spaces giving rise to a non-trivial geometry. Ultimately using the deformed geometry we have calculated the geodesic distance between two points in appropriate subspaces and have interpreted them as the quadratic parts of deformed Casimirs of the respective algebras to write down a deformed dispersion relation. Finally, we have concluded with some remarks in section-4.  
	\section{Revisiting Poincare algebra and its various contractions}
	The Poincare group ($\text{ISO}(1,3)=  \text{SO}(1,3)\ltimes \mathcal{J}^4 $, with $\mathcal{J}^4 \approx \mathbb{R}^{1,3}$ being the additive group of space-time translations) is the isometry group of flat Minkowski space-time ($\mathcal{M}$) which can be described by the Cartesian coordinates $\{q^{\mu}\}$ with associated metric $\eta$ in its mostly negative convention given by,
	\begin{align}
		\eta &:= \left(dq^0\otimes dq^0
		- \delta_{ij} dq^i\otimes dq^j
		\right)=\eta_{\mu\nu}\,\,dq^{\mu}\otimes dq^{\nu}\label{1}
		\intertext{and its inverse is,}
		\eta^{-1}
		&= \left(\deldel{q}{}{0}\otimes \deldel{q}{}{0}
		- \delta^{ij} \deldel{q}{}{i}\otimes \deldel{q}{}{j}
		\right)=\eta^{\mu\nu}\,\frac{\partial}{\partial q^{\mu}}\,\otimes\,\,\frac{\partial}{\partial q^{\nu}}\label{2}
	\end{align}
	The generators of the Poincare Lie algebra $\mathfrak{iso}(1,3)$ corresponding to boost, rotations and 
	translations in space and time can be written in terms of vector fields on $\mathcal{M}$ as,
	\begin{align}
		J_{\mu\nu}
		&= 
		\left( 
		q_\mu\deldel{q}{}{\nu} - q_\nu\deldel{q}{}{\mu}
		\right);\qquad    P_\mu = \deldel{q}{}{\mu}\label{representation}
	\end{align}
	fulfilling
	\begin{align}
		[J_{\mu\nu},J_{\rho\sigma}]&=(\eta_{\nu\rho}J_{\mu\sigma}+\eta_{\mu\sigma}J_{\nu\rho}-\eta_{\mu\rho}J_{\nu\sigma}-\eta_{\nu\sigma}J_{\mu\rho})\nonumber\\
		[J_{\mu\nu},P_{\rho}]&=(\eta_{\nu\rho}P_{\mu}-\eta_{\mu\rho}P_{\nu})\nonumber\\
		[P_{\mu},P_{\nu}]&=0\label{poincarealgebra}
	\end{align}
	It should be noted that the representations \eqref{representation}
	of the classical Poincare algebra of \eqref{poincarealgebra} stems from its representation furnished by the space $C^{\infty}(\mathcal{M})$ of scalar fields $\phi(q):=\phi(\vec{q},t)$ on space-time. These scalar fields $\phi(q)$, under the action of Poincare group transform, by definition, as
	\begin{equation}
		\phi(q)\,\,\to\,\,\phi^{\Lambda}(q)=\phi(\Lambda^{-1}q);\qquad \Lambda\,\,\in\,\, \text{ISO}(1,3)\label{A1}
	\end{equation}
	To recall, how these particular form arises, one just need to consider infinitesimal Poincare transformation. For an explicit illustration, let us take $\Lambda$ to be an element, very close to the identity, of the homogeneous Lorentz group:
	\begin{equation}
		\Lambda =I+\omega\quad \in\,\,\text{SO}(1,3);\qquad \omega^T=-\omega \,\,\,\textrm{with}\,\,\, |\omega^{\mu}\,_{\nu}|\ll 1\label{A2}
	\end{equation}
	One can then show that the functional increment induced in $\phi(q)$ i.e. the so called Lie-increment is given by 
	\begin{equation}
		\delta_0\phi(q):=\phi^{\left(I+\omega\right)}(q)-\phi(q)=\phi((I-\omega)q)-\phi(q)=\frac{1}{2}\omega^{\mu\nu}J_{\mu\nu}\phi(q)\label{A3}
	\end{equation}
	with $J_{\mu\nu}$ given precisely by \eqref{representation}. One can like-wise, obtain the representation of $P_{\mu}$ in \eqref{representation} by considering an infinitesimal translation $q^{\mu}\,\,\to\,\,q'^{\mu}=q^{\mu}+a^{\mu};\quad |a^{\mu}|\ll 1$.
	The corresponding quantum Poincare algebra generators are obtained by simply multiplying the generators in \eqref{representation} by $(-i)$: 
	\begin{align*}
		\numberthis
		\label{qrepresentation}
		\begin{split}
			\widehat{J}_{\mu\nu}
			&= -i J_{\mu\nu},\qquad
			\widehat{P}_\mu= -i P_\mu
		\end{split}
	\end{align*}
	fulfilling the following quantum Poincare Lie algebra:
	\begin{align}
		\comm{\widehat{J}_{\mu\nu}}{\widehat{J}_{\rho\sigma}}
		&=-i\left(\eta_{\nu\rho}\widehat{J}_{\mu\sigma}+\eta_{\mu\sigma}\widehat{J}_{\nu\rho}-\eta_{\mu\rho}\widehat{J}_{\nu\sigma}-\eta_{\nu\sigma}\widehat{J}_{\mu\rho}\right) \nonumber
		\\
		\comm{\widehat{J}_{\mu\nu}}{\widehat{P}_{\rho}}&=-i \left( \eta_{\nu\rho}\widehat{P}_{\mu}-\eta_{\mu\rho}\widehat{P}_{\nu}\right) \nonumber 
		\\
		\label{poincarealgebra2}
		\comm{\widehat{P}_{\mu}}{\widehat{P}_{\nu}}&=0
	\end{align}
	As we know,	Einstein introduced the coordinate 
	$q^0=ct$ to unify space-time on a consistent framework, i.e. to endow them with the same dimension of length,  where $t$ represents relativistic time. It's important to note that in this paper, we exclusively use $t=\frac{q^0}{c}$ to signify relativistic time and, regard it as a function of $c$ : $t(c)$, enabling us to explore two following limits:  $c\to\infty$ (non-relativistic or Galilean limit) and $c\to 0$ (ultra-relativistic or the so called Carrollian limit) 
	resulting in non-relativistic ($q^0_g$) and ultra-relativistic $(q_c^0)$ time which are defined as $q_g^0 = \displaystyle{\lim_{c\to \infty}
		t(c)} $ and $q_c^0 = \displaystyle{\lim_{c\to 0} t(c)} $ respectively.  
	\subsection{Non-relativistic limit of Poincare algebra : Bargmann Algebra}
	The Galilean time and space coordinates $({q_g^0},q_g^i)$ are obtained by taking the limit $c\to \infty$ of the Minkowski space coordinates $(q^\mu)$. However, it should be noted that the contraction procedure employed below differs slightly in approach from the one adopted in   \cite{inonu1953contraction}. To this end, consider the transformation
	\begin{align*}
		\label{gcoordinates}
		\numberthis
		\begin{split}
			q^0 &\longrightarrow q_g^0 = \lim_{c\to\infty} t(c) 
			= \lim_{c\to\infty} 
			\left(\frac{q^0}{c}\right);\qquad		q^i \longrightarrow q_g^i =  q^i
		\end{split}
	\end{align*}
	The corresponding inverse space-time metric $\gamma^{-1}$ for the flat Newton-Cartan space-time $\mathcal{N}$ can be shown to be derived from inverse of the flat Minkowski metric $\eta^{-1}$ \eqref{2} to get its degenerate form as follows,
	\begin{align}
		\gamma^{-1} := \lim_{c\to\infty}\eta^{-1}(q^0,q^i;c)
		=\lim_{c\to\infty}\left( \frac{1}{c^2}\deldel{t}{}{}
		\otimes\deldel{t}{}{}
		-
		\delta^{ij}\deldel{q}{}{i}\otimes\deldel{q}{}{j}
		\right)
		= -
		\delta^{ij}\deldel{q}{g}{i}\otimes\deldel{q}{g}{j}\label{ncmetric}
	\end{align}
	Clearly this cannot be inverted to get $\gamma$, consequently although $\gamma^{-1}$ can be used to raise spatial indices to obtain the covariant couterpart $q_i^g$ of $q_g^i$ \eqref{gcoordinates} as,
	\begin{equation}
		q_g^i = \gamma^{i\mu}q_\mu^g = -\delta^{ij}q_j^g = -q_i^g   
	\end{equation}
	the same cannot be done for temporal coordinate. In other words, we do not have $q^g_0$ here. \\  \\  
	Having defined the Galilean space-time coordinates we now try to derive the functional representations for the various generators of the Bargmann algebra. For that, we need to implement the Galilean reduction of $C^{\infty}(\mathcal{M})$ introduced above, on which the action of relativistic Hamiltonian of a free particle can be defined by identifying the Hamiltonian operator as $\widehat{H}= -\widehat{P}_t$ where $\widehat{P}_t=-i\deldel{t}{}{}$ (see \eqref{qrepresentation})   \cite{weinberg_qft1}. Using the relativistic dispersion relation, one obtains, by using $\widehat{H}= \left(\widehat{P}_i^2c^2
	+ m^2c^4 \right)^{\sfrac{1}{2}}$,
	the well known Klein-Gordon equation with $\phi(q)$ satisfying \eqref{A1}.
	\begin{equation}
		\label{eq4}
		\left(\frac{1}{c^2}\frac{\partial^2}{\partial t^2}-\vec{\nabla}^2+m^2c^2\right)\phi(q)=0; \myquad[2]
		\phi(\vec{q},t)\,\in\,C^{\infty}(\mathcal{M})
	\end{equation} 
	Now to extract the effective non-relativistic limit of the above expression we introduce another field $\psi(\vec{q},t)$ by defining   \cite{Chakraborty:2006hv},
	\begin{equation}
		\label{eq3}
		\psi(\vec{q},t):=e^{imc^2t} \phi(\vec{q},t)
	\end{equation}
	With this we have essentially extracted out the rest mass energy contribution from the phase and the resulting $\psi(\Vec{q},t)$ no longer transforms as a scalar under the entire Poincare group. As is well known $\psi(\Vec{q},t)$ can be identified as the Sch\"{o}dinger's field i.e. satisfies Schrodinger equation   \cite{Levy-Leblond:1967eic,Bergshoeff:2022eog} in the limit $c \to\,\infty: \,\,\psi(\vec{q},t)\to \psi(\vec{q}_g,q^0_g)\,\in\,C^{\infty}(\mathcal{N})$, where only positive frequency components survives and negative energy components drops out\footnotemark . 
	\footnotetext{Strictly speaking, a prefactor $\frac{1}{\sqrt{2mc}}$ is required in the R.H.S. of \eqref{eq3} to implement the non-relativistic reduction, where for example the Klein-Gordon action for a free complex scalar field goes over to the action of a free Schr\"{o}dinger field. But we are suppressing this factor here, as it has no bearing on the subsequent computations presented below.}
	\\\\
	To start with, let us spell out the strategy for the construction of the various Poincare generators in the limit $c\to\infty$ to obtain the corresponding Galilean generators. For that we'll express the functional representations of the corresponding Poincare generators in terms of the Galilean coordinates \eqref{gcoordinates}. This substitution will introduce certain factors of $c$ in the respective expressions. Now, to ensure the representation converges in the limit $c \to \infty$, we must re-scale the generators by a factor of $c$ or $\frac{1}{c}$, as appropriate. The convergent form of these re-scaled generators will then correspond to the final representation for the Galilean generators. Naturally, these representations will deviate from the standard Poincare ones, particularly for the Boost ($\widehat{J}_{i0}$) and energy ($\widehat{P}_0$), as these expressions contains derivatives with respect to the temporal coordinate, and note in particular, that it was the time variable only which was re-scaled in \eqref{gcoordinates} and the extra phase in \eqref{eq3} has only $t$-dependence.
	\begin{itemize}
		\item \textbf{Galilean boost:} We therefore start with the representation of Poincare boost generator as follows,
		\begin{align*}
			\widehat{J}_{i0}&= -i \left(q_i\deldel{q}{}{0} - 
			q_0 \deldel{q}{}{i}\right)=-i\left(q_i\deldel{q}{}{0} - q^0 \deldel{q}{}{i}\right)= -i\left(\frac{q_i}{c}\deldel{t}{}{} 
			- c t \deldel{q}{}{i}\right)
			\intertext{Now scaling this by $\sfrac{1}{c}$, let us first introduce}
			\numberthis\label{B1}
			\widehat{\bar{B}}_i &:= 
			\left(\frac{\widehat{J}_{i0}}{c}\right) 
			= 
			-iq_i^g 
			\left(\frac{1}{c^2}\deldel{t}{}{}\right)
			+ i\left(t(c) \deldel{q}{g}{i}\right)
		\end{align*} 
		\noindent
		Here, it is essential to note that we haven't taken $c\to\infty$ limit as yet\footnotemark. As we show in the sequel that the object $\widehat{\bar{B}}_i$ needs to first undergo a suitable unitary transformation before taking $c\to\infty$ limit. This stems from the fact that the operator $\frac{1}{c^2}\deldel{t}{}{}$
		in the limit, $c\to\infty$ does not necessarily vanish when acting on an arbitrary $\phi(\vec{q},t)\,\,\in\,\,C^{\infty}(\mathcal{M})$ fulfilling \eqref{eq4}.  
		
		\footnotetext{Note that once the limit is taken, the resulting expression will no longer be invertible! This is because once the limit is taken there is no way to retrieve the parent form of the Poincare generator. This will be a recurrent feature in our constructions of various other generators in the respective limiting cases}
		To see this, just take the plane wave solution \eqref{eq4} 			
		\begin{equation}
			\label{A5}
			\phi(q)=e^{-i(\omega_pt-\vec{p}\cdot\vec{q})} \quad\textrm{with}\qquad\omega_p=\sqrt{\overrightarrow{p}^2c^2+m^2c^4}=mc^2+\frac{\overrightarrow{p}^2}{2m}+\mathcal{O}\left(\frac{1}{c^2}\right)
		\end{equation}
		It then readily follows that,
		\begin{align}
			\label{A6a}
			\lim_{c\to\infty}\left|\frac{1}{c^2}
			\frac{\del \phi(\vec{q},t)}{\del t}\right|
			= m\neq 0
		\end{align}
		On the other hand, if we consider the action of
		$\frac{1}{c^2}\deldel{t}{}{}$ on the field
		$\psi(\vec{q},t)$ \eqref{eq3} and take the limit $c\to\infty$,
		then we get on using \eqref{A5},
		\begin{align}
			\label{A6b}
			\lim_{c\to\infty}\left|\left(\frac{1}{c^2}
			\frac{\del \psi(\vec{q},t)}{\del t}\right)\right|
			= \lim_{c\to\infty}
			\frac{1}{c^2}\left(\omega_p - mc^2\right)
			= 0
		\end{align}
		This is expected as we have dispensed with the rest 
		energy term. This yields, on using \eqref{A5}, in
		the Galilean limit the well known non-relativistic dispersion relation.
		\begin{align}
			\label{A7}
			E_p = \lim_{c\to\infty}
			\left(\omega_p - mc^2\right)
			= \frac{\overrightarrow{p}^2}{2m}
		\end{align}
		Furthermore, it is well-known that Wigner-In\"{o}nu 
		group contraction of inhomogenous Lorentz group 
		ISO(1,3) in the limit $c\to\infty$ gives way to 
		inhomogenous Galileo group, where only the boost transformation gets deformed to
		\begin{align}
			\label{A8}
			q_g^0 \longrightarrow {q'}_g^0
			= q_g^0 \quad \text{and} \quad
			q_g^i \longrightarrow {q'}_g^i
			= \left(q_g^i - v^i q_g^0\right)
		\end{align}
		where, $v^i$ is the relative velocity connecting
		these two frames: unprimed and primed frames and,
		is related to $(\omega^{i0})$ which is actually the relativistic rapidity parameter along the $i^{\text{th}}$ direction and is related to the velocity parameter $v^i$ as $\omega^{i0}=\frac{v^i}{c}$ in the limit $c\to\infty$. And, the space $C^\infty(\mathcal{N})$ furnishes only a projective representation   \cite{weinberg_qft1,Bargmann:1954gh,Chakraborty:2006hv} of the Galilean group. It thus becomes clear that to find a representation of the Galileo boost \eqref{A8} in  $C^\infty(\mathcal{N})$, starting from the expression of Lorentz boost, we need to compute the \emph{Lie increment} $\delta_0 \psi(\Vec{q}_g,q_g^0)$ rather than $\delta_0 \phi(\vec{q},t)$\eqref{A3}. To that end, consider a Lorentz boost along $i$-th direction so that only $\omega^{i0}=-\omega^{0i}\neq 0$ in \eqref{A3}. So it reduces to
		%%%%%%%%%%%
		\begin{align}
			\label{A9}
			\delta_0\phi(\vec{q},t) = \omega^{i0}
			J_{i0}\phi(\vec{q},t)
		\end{align}
		Now it follows trivially from \eqref{A3} that
		$\delta_0\phi(\Vec{q},t)
		= e^{-imc^2t}\delta_0\psi(\Vec{q},t)$. We can therefore recast \eqref{A9} as,
		\begin{align}
			\label{A10}
			\delta_0\psi(\Vec{q},t)
			= \omega^{i0}\left(e^{imc^2t}\widehat{J}_{i0}
			e^{-imc^2t}\right)\psi(\Vec{q},t)
		\end{align}
		Finally, we introduce the Galileo boost 
		generator $\widehat{B}_i^g$, using
		\eqref{B1} as,
		\begin{align}
			\label{A11}
			\widehat{B}_i^g
			:= \lim_{c\to\infty}
			\left( e^{imc^2t}\widehat{\bar{B}}_i
			e^{-imc^2t}\right)
		\end{align}
		where we have first subjected the scaled $\widehat{J}_{i0}$
		by a suitable unitary transformation and then executing
		the limit. One can then make use of 
		the simple identity: $e^{imc^2t}\del_t
		e^{-imc^2t}= \left(\del_t - imc^2\right)$
		and the fact that $\del_0=\deldel{q}{0}{}
		= \frac{1}{c}\del_t$ to obtain the final form of
		Galilean boost as,
		\begin{align}
			\label{rev_m_gal_eq4a}
			\widehat{B}_i^g
			= 
			\left(
			iq_g^0\deldel{q}{g}{i} - mq_i^g
			\right)
			=-\left(
			q_g^0\widehat{P}_i^g + mq_i^g
			\right)
		\end{align}
		acting on $\psi(q_g)=\psi(\vec{q}_g,q_g^0)\in
		C^\infty(\mathcal{N})$.
		Clearly, this does not have the form of a vector field anymore  due to the contribution of the second term involving mass. In fact the central extension of the Galilean algebra   \cite{Bargmann:1954gh,weinberg_qft1} owes its origin to this mass term. 
		\item 
		\textbf{Euclidean Subalgebra} $\left(\mathfrak{iso}(3)=\text{Lie}\left(\text{SO}(3)\ltimes \mathcal{J}^3\right)\right)$\textbf{:} The representation of the generators of this sub-algebra will remain the same as the expression only involves spatial coordinates which remains un-scaled (see \eqref{gcoordinates}):
		\begin{subequations}
			\begin{align}
				\label{rev_m_gal_eq4b}
				\widehat{J}_{ij}^g &= \widehat{J}_{ij}=
				-i\left(q_i^g\deldel{q}{g}{j} - q_j^g \deldel{q}{g}{i}\right)
				= \left(q_i^g\widehat{P}_j^g - q_j^g \widehat{P}_i^g\right)
				\in \mathfrak{so}(3);
				\\
				\label{rev_m_gal_eq4c}
				\widehat{P}_i^g &= \widehat{P}_i
				=-i\deldel{q}{}{i}  = -i\deldel{q}{g}{i} 
				= \widehat{P}_i^g
				\in \mathfrak{J}^3
			\end{align}
		\end{subequations}
		\item \textbf{Galilean time translation:} Like the boost generators $\widehat{J}_{i0}$, the Poincare time translation generator\footnotemark, given by 
		%%%%%%%%
		\footnotetext{It is evident to note that the relativistic time translation operator $\widehat{P}_t$ is actually the negative of the relativistic Hamiltonian($\widehat{H}$) i.e. $\widehat{H}= -\widehat{P}_t$ where $\widehat{P}_t=-i\deldel{t}{}{}$ (see \eqref{qrepresentation}).}
		%%%%%%%%
		\begin{align*}
			\numberthis
			\widehat{P}_0 &= -i\deldel{q}{}{0}
			= -i\left(\frac{1}{c}\deldel{t}{}{}\right)
			:= \frac{1}{c}\widehat{P}_t \, ,
			\intertext{also needs to undergo the same unitary transformation, as we make transition from the domain of relativistic field $\phi(q)$ to $\psi(\Vec{q}_g,q_g^0)\in C^{\infty}(\mathcal{N})$- the space of Schr\"{o}dinger fields as}
			\numberthis
			\label{rev_m_gal_eq4d}
			\widehat{\Tilde{P}}_0
			&=e^{imc^2t}\widehat{P}_0 e^{-imc^2t}=\frac{1}{c}
			\left(\widehat{P}_t-mc^2\right)
		\end{align*}
		Finally, we scale it by $c$ to define Galilean translational generator as, 
		\begin{equation}
			\label{q8}
			\widehat{P}_0^g:= \lim_{c\to\infty}\,c  \widehat{\tilde{P}}_0=\lim_{c\to\infty}
			\left(\widehat{P}_t-mc^2\right)
		\end{equation}
		Although both $P_t$ and $mc^2$ diverges in the limit $c\to\infty$, their difference converges. For free particle described by the plane wave \eqref{A5} we have seen that difference indeed gives finite energy $E_p$ in this limit \eqref{A7}. However, the limit should be taken after computing the whole set of commutators below: 
		\begin{align*}
			\comm{\widehat{J}_{ij}^g}{\widehat{J}_{rs}^g} 
			&= -i \left(-\delta_{ir}\widehat{J}_{js}^g 
			+ \delta_{jr}\widehat{J}_{is}^g + \delta_{is} 
			\widehat{J}_{jr}^g
			- \delta_{js}\widehat{J}_{ir}^g\right),
			&\comm{\widehat{J}_{ij}^g}{\widehat{B}_r^g} &= 
			-i \left(\delta_{jr}\widehat{B}_i^g - \delta_{ir}
			\widehat{B}_j^g\right),
			\\
			\comm{\widehat{J}_{ij}^g}{\widehat{P}_0^g}&= 0
			&\comm{\widehat{J}_{ij}^g}{\widehat{P}_r^g} &= 
			-i \left(\delta_{jr}\widehat{P}_i^g - \delta_{ir}
			\widehat{P}_j^g \right),
			\\
			\comm{\widehat{P}_i^g}{\widehat{P}_j^g} 
			&= 0=\comm{\widehat{P}_0^g}{\widehat{P}_i^g},
			&\comm{\widehat{B}_i^g}{\widehat{B}_j^g} &= 0
			\\
			\numberthis
			\label{mgal}
			\comm{\widehat{B}_i^g}{\widehat{P}_j^g}  
			&= im\delta_{ij},
			&\comm{\widehat{P}_0^g}{\widehat{B}_i^g} 
			&= i \widehat{P}_i^g
		\end{align*}
	\end{itemize}
	This is the final algebra with the mass `$m$' being the central extension famously known as Bargmann algebra as introduced in   \cite{Bargmann:1954gh}. It should be clear at this stage that, like the original $\widehat{P}_t$, $\widehat{P}_0^g$ in \eqref{q8} too can be represented as $\left(-i\deldel{q}{g}{0}\right)$, provided its domain of action is on the space of functions like $\psi(\vec{q}_g,q_g^0)\in C^{\infty}(\mathcal{N})$. This is clear from the non-relativistic dispersion relation \eqref{A7}, as the corresponding plane wave will be of the form $e^{-i(E_pt-\vec{p}\cdot\vec{q})}$.  
	And clearly, the massless limit cannot be taken
	by simply putting $m=0$ in the massive theory. \\
	The above algebra has three Casimirs   \cite{Figueroa-OFarrill:2024ocf, Bacry:1968zf}: the central extension or the `Mass' generator $\widehat{\mathfrak{C}}_1=m$ and the quadratic Casimirs given by
	%%%%%%%%%%%%%%%%
	\begin{align}
		\widehat{\mathfrak{C}}_2 &= \norm{m\widehat{\overrightarrow{J}}^g+\widehat{\overrightarrow{P}}^g\times\widehat{\overrightarrow{B}}^g}^2; \quad\textrm{where}\,\,\,\widehat{J}^g_i=\frac{1}{2}\epsilon_{ijk}\widehat{J}_{ij}^g\label{gcasimir1}\\
		\widehat{\mathfrak{C}}_3
		&=\left(2m\widehat{P}_0^g + \widehat{\overrightarrow{P}}_g^2\right)
		\label{gcasimir}
	\end{align}
	$\widehat{\mathfrak{C}}_2$ \eqref{gcasimir1} can be recognized as the Galilean counterpart of the square of Pauli-Lubanski vector and $\widehat{\mathfrak{C}}_3$ given by \eqref{gcasimir}, can be recognized as mass-shell condition on which we would like to make some pertinent observations. Firstly, note that this form of Casimir \eqref{gcasimir} is not completely quadratic in all the components of energy and (spatial) momenta; the quadratic parts of the Casimir involve only the spatial components. It immediately follows that 3-momentum space only inherits the flat metric after this contraction. We have more to say on this in the sequel. It is evident to note that the Galilean energy operator is given as $\widehat{H}_g = \left(i\deldel{q}{g}{0}\right) = -\widehat{P}_0^g$, which appears as the Hamiltonian operator in the well-known Schr\"{o}dinger's equation. From this we can again identify the Galilean dispersion relation as $E_p=\frac{\vec{P}_g^2}{2m}$ \eqref{A7}  for a free non-relativistic particle, with $E_p$ being the eigen value of $\hat{H}_g$.
	\subsection{Galilean Algebra}
	Although normally one cannot have massless particle moving in the Newton-Cartan space-time, one can nevertheless introduce the Galilean algebra as shown in   \cite{axel_non_lorentzian,townsend_tachyons}, by considering non-relativistic ($c\to\infty$) limit of a \textit{tachyonic} particle. To recapitulate this in this section, we shall derive the corresponding dispersion relation for massless non-relativistic particles. For that, we start with the relativistic dispersion relation re-written as
	\begin{align}
		\left(\overrightarrow{p}^2 + m^2 c^2\right)
		&= \frac{E^2}{c^2}
		\intertext{Clearly a meaningful limit $c\to\infty$ can be taken iff we take the limit $m \to 0$ simultaneously, holding $E$ fixed, so that the product $mc$ converges to some imaginary value yielding	}
		\numberthis
		\label{eq1}
		\left|\overrightarrow{p}\right|
		&= 
		i m c
	\end{align}
	In other words, the particle satisfying \eqref{eq1} must be a tachyon to begin with, i.e.
	possessing imaginary mass and moving with a speed
	faster than light
	\cite{sudarshan1969nature,dhar1968quantum}. Equivalently, by replacing  $im\to m$ we have
	\begin{align*}
		\numberthis
		\label{eq2}
		\left|\overrightarrow{p}\right|
		&= 
		m c
	\end{align*}
	where $m$ is now an imaginary number and should not be identified with mass anymore. Equation \eqref{eq2} implies $\overrightarrow{p}^2 = m^2c^2$, where $m\to 0$ and $c\to\infty$, ensuring that the product $mc$ remains finite and becomes associated with the concept of `color'   \cite{souriau2012structure,axel_non_lorentzian}—a quantity that remains constant for a given particle.\\
	Now we revisit the construction of the boost generator in our own approach- as the strategy for which has already been spelled out in the previous section.  
	\begin{itemize}
		\item \textbf{Galilean boost:} Since, $m=0$
		in thus case, the necessary unitary transformation occurring in \eqref{A11} is no longer required and the representation of the boost is simply given by.
		\begin{equation}
			\widehat{B}_i^g 
			:= \lim_{c\to \infty}\left(\frac{\widehat{J}_{i0}}{c}\right)
			=
			-i	q_i^g \lim_{c\to\infty}
			\left(\frac{1}{c^2}\deldel{t}{}{}\right)
			+i \left( \lim_{c\to\infty}t(c) \right)\deldel{q}{g}{i}
		\end{equation}
		However, for $~m=0,\quad \frac{1}{c^2}\deldel{t}{}{}\to 0$ as $c\to\infty$ even if we consider its action on $C^{\infty}(\mathcal{M})$ (putting simply $m=0$ in  \eqref{A6a}). So the representation finally reduces to
		\begin{equation}
			\label{rev_ml_gal_eq1a}
			\widehat{B}_i^g= i q^0_g \deldel{q}{g}{i}
		\end{equation} 
		One can also check that this generator indeed produces the following transformations in the coordinate sector:
		\begin{equation}
			q_g^0\,\,\longrightarrow\,\,     {q'}_g^0 = q_g^0, \qquad
			q_g^i\,\,\longrightarrow\,\,	{q'}_g^i = q_g^i - v^i q_g^0
		\end{equation}
		where $v_i$ is the boost parameter in the respective spatial direction.\\
		The rotation $\left(\widehat{J}_{ij}\right)$ and translation generators $\left(\widehat{P}_t, \widehat{P}_i\right)$ remain the same as that of the massive case discussed in the previous section. There exist one contrasting feature with the Bargmann algebra generators though: here in the limit $c\to \infty$ the zeroth component of momentum $\widehat{P}_t$ is exactly equal to $\widehat{P}_0^g$ (see \eqref{q8} in the case $m=0$). 
	\end{itemize}
	The Galilean algebra $\mathfrak{g} (d,1)$ can now be derived as,
	%%%%%%%%%%%
	\begin{align*}
		\comm{\widehat{J}_{ij}^g}{\widehat{J}_{rs}^g} &=-i \left(-\delta_{ir}\widehat{J}_{js}^g 
		+ \delta_{jr}\widehat{J}_{is}^g + \delta_{is} \widehat{J}_{jr}^g
		- \delta_{js}\widehat{J}_{ir}^g\right),
		&\comm{\widehat{J}_{ij}^g}{\widehat{B}_r^g} &= -i
		\left(\delta_{jr}\widehat{B}_i^g - \delta_{ir}\widehat{B}_j^g\right),
		\\
		\comm{\widehat{J}_{ij}^g}{\widehat{P}_0^g }&= 0
		&\comm{\widehat{J}_{ij}^g}{\widehat{P}_r^g} &= -i
		\left(\delta_{jr}\widehat{P}_i^g - \delta_{ir}\widehat{P}_j^g \right),
		\\
		\comm{\widehat{P}_i^g}{\widehat{P}_j^g} &= 0=\comm{\widehat{P}_0^g}{\widehat{P}_i^g},
		&\comm{\widehat{B}_i^g}{\widehat{B}_j^g} &= 0
		\\
		\numberthis
		\label{galalg}
		\comm{\widehat{B}_i^g}{\widehat{P}_j^g} &= {\color{blue} 0},
		&\comm{\widehat{P}_0^g}{\widehat{B}_i^g} &= i\widehat{P}_i^g
	\end{align*}
	Note that
	\begin{equation}
		\left(\widehat{P}_i^g\right)^2=\widehat{P}_i^g \widehat{P}_i^g=\left(\widehat{\overrightarrow{P}}\,^g\right)^2\label{a1}
	\end{equation}
	can be recognized as one of the Casimir of the above algebra. Note that, in contrast to the corresponding expression of the Bargmann algrebra \eqref{gcasimir}, this expression of the Casimir operator is entirely quadratic, in the spatial components of momentum and involves no energy term. So again, it is a 3-momentum space, which inherits the flat Euclidean metric. 
	\subsection{Ultra-relativistic limit of Poincare algebra: Carroll Algebra}
	Like in the non-relativistic case here too we can define the Carrollian coordinates as the formal $c\to  0$ limit of Minkowski ones   \cite{inonu1953contraction} as,
	\begin{align*}
		\numberthis
		\label{carroll}
		\begin{split}
			q^0 \longrightarrow
			{q_c^0} = \lim_{c\to 0} t(c)
			=\lim_{c\to 0}\left(\frac{q^0}{c}\right)
			\, ;
			\myquad[2]
			q^i \longrightarrow
			q_c^i = q^i
		\end{split}
	\end{align*}
	The flat Carrollian metric is defined as,
	\begin{align}
		\label{C1}
		\cg = \lim_{c\to 0}\eta(q^0,q^i;c)
		=\lim_{c\to 0} 
		\left(c^2 dt \otimes dt	- \delta_{ij} dq^i \otimes dq^j\right)= - \delta_{ij} dq_c^i \otimes dq_c^j
	\end{align}
	Note that unlike the Newton-Cartan metric $\gamma^{-1}$ \eqref{ncmetric} this metric  $\cg$ in its covariant form becomes \emph{degenerate}. Accordingly, this $\cg$, can be used to lower the spatial components \emph{only} as,
	\begin{align*}
		\numberthis
		\begin{split}
			q_i^c & =\cg_{i\mu}q_c^\mu= -\delta_{ij}q_c^j = -q_c^i
		\end{split}
	\end{align*}
	And, because of the degeneracy of $\cg$ here too there is no concept of covariant temporal coordinate ($q_0^c$), just like the Galilean case, albeit for different reasons.\footnote{In the Galilean case it diverges and in the Carrollian case it goes to zero.} 
	\\ 
	Next we shall find out the Carrollian generators which are the generators of the group of isometry of the Carroll manifold($\mathcal{C}$).
	\begin{itemize}
		\item \textbf{Carrollian boost:} We will adopt a similar methodology as in the Galilean case to determine the corresponding generators. Initially, by expressing the Poincaré boost generators in terms of the Carrollian coordinates, we obtain
		\begin{align*}
			\widehat{J}_{i0} &=-i \left(q_i\deldel{q}{}{0} - 
			q_0 \deldel{q}{}{i}\right)
			=-i \left(q_i\deldel{q}{}{0} - 
			q^0 \deldel{q}{}{i}\right)
			=-i \left(\frac{q_i^c}{c}\deldel{t}{}{}
			- c t\deldel{q}{c}{i}\right)
			\intertext{Now for the limit $c \to\,0$ to exist we define the Carrollian boost generator 
				as following}
			\numberthis
			\label{rev_carr_alg_eq5c}
			\widehat{B}_i^c &:= \lim_{c\to 0} c\widehat{J}_{i0}
			= -i\left(q_i^c\deldel{q}{c}{0}\right)
		\end{align*}
		One can just check that this will generate the following transformation in the coordinate sector
		\begin{equation}
			q_c^0 \,\,\longrightarrow\,\, {q'}_c^0 = \left({q_c^0} + \beta_i q_c^i\right), \myquad[3]	q_c^j \,\,\longrightarrow\,\,{q'}_c^j = q_c^j 
		\end{equation}
		where $\beta_i$ are the boost parameters. Note that under the transformations of Carrollian boost does not alter spatial coordinates; rather, it exclusively affects the temporal coordinate and, in some sense, is dual  \cite{duval_carroll} to the Galilei boost \eqref{A8}. 
		\item \textbf{Carrollian $\mathfrak{iso}(3)$:} The $\mathfrak{iso}(3)$ generator will be same as that of the Minkowski counterparts as they do involve no temporal derivatives. 
		\begin{subequations}
			\begin{align*}
				\numberthis
				\label{rev_carr_alg_eq5d}
				\widehat{J}_{ij}^c &= \widehat{J}_{ij}
				=-i \left(q_i^c\deldel{q}{c}{j} - q_j^c \deldel{q}{c}{i}\right)
				= \left(q_i^c \widehat{P}_j^c
				- q_j^c \widehat{P}_i^c
				\right)\in \mathfrak{so}(3);
				\\
				\numberthis
				\label{rev_carr_alg_eq5b}
				\widehat{P}_i^c &=  \widehat{P}_i= -i\deldel{q}{}{i}
				= -i\deldel{q}{c}{i}
				\in \mathfrak{J}^3
			\end{align*}
		\end{subequations}
		\item \textbf{Carrollian time translation: } Rewriting the usual time translation generator in terms of Carrollian coordinates we define
		Carrolian time translation as
		\begin{align*}
			\numberthis
			\label{rev_carr_alg_eq5a}
			\widehat{P}_0^c &:= \lim_{c\to 0} c \widehat{P}_0 
			= -i\deldel{q}{c}{0}
		\end{align*}
	\end{itemize}
	Using the above representation one can verify the following Lie algebra   \cite{SenGupta:1966qer} between the generators 
	\begin{align*}
		\comm{\widehat{J}^c_{ij}}{\widehat{J}^c_{rs}} & =-i \left(-\delta_{ir}\widehat{J}^c_{js} 
		+ \delta_{jr}\widehat{J}^c_{is} + \delta_{is} \widehat{J}^c_{jr}
		- \delta_{js}\widehat{J}^c_{ir}\right),
		&\comm{\widehat{J}^c_{ij}}{\widehat{B}^c_r}&= -i
		\left(\delta_{jr}\widehat{B}^c_i-\delta_{ir}\widehat{B}^c_j\right),
		&\comm{\widehat{J}^c_{ij}}{\widehat{P}^c_0}&= 0
		\\
		\comm{\widehat{J}^c_{ij}}{\widehat{P}^c_r}&= -i
		\left(\delta_{jr}\widehat{P}^c_i-\delta_{ir}\widehat{P}^c_j\right),
		&\comm{\widehat{P}^c_i}{\widehat{P}^c_j}&= 0=\comm{\widehat{P}^c_0}{\widehat{P}^c_i},
		&\comm{\widehat{B}^c_i}{\widehat{B}^c_j}&= 0
		\\
		\comm{\widehat{B}^c_i}{\widehat{P}^c_j}&= i \left(\delta_{ij} \widehat{P}^c_0\right) ,
		& \comm{\widehat{P}^c_0}{\widehat{B}^c_i} &= 0
		\\
		\numberthis
		\label{rev_carr_alg_eq2}
	\end{align*}
	Interestingly, if we formally identify the Carrollian boost generator $\widehat{B}_i^c$ as $\widehat{J}_{i0}^c$ then, the Carroll algebra \eqref{rev_carr_alg_eq2} can also be written in the following covariant form,
	%%%%%%%%%%%%%%%%%%%%%%%%
	\begin{align*}
		\numberthis
		\label{rev_carr_alg_eq3a}
		\myquad[2]
		\begin{split}
			&\comm{\widehat{J}^c_{\rho\sigma}}{\widehat{J}^c_{\mu\nu}}
			= -i\left(\cg_{\rho \mu}\widehat{J}^c_{\sigma\nu}
			- \cg_{\sigma \mu}\widehat{J}^c_{\rho\nu}
			- \cg_{\rho\nu}\widehat{J}^c_{\sigma\mu}
			+ \cg_{\sigma\nu}\widehat{J}^c_{\rho\mu}\right)
			\\
			&\comm{\widehat{J}^c_{\rho\sigma}}{\widehat{P}^c_\mu}
			\hspace{1mm}
			=-i \left(\cg_{\rho\mu}\widehat{P}^c_\sigma
			- \cg_{\sigma\mu}\widehat{P}^c_\rho\right)
		\end{split}
	\end{align*}
	\noindent
	where $\cg$ is the \emph{degenerate}
	Carrollian metric\eqref{C1} given by $g=\text{diag} (0,-1,-1,-1)$.\\
	There are two Casimirs of the algebra   \cite{levy_caroll} given by can be shown to be given by 
	\begin{equation}
		\widehat{\mathfrak{I}}_1=\widehat{P}_0^c;\qquad \widehat{\mathfrak{I}}_2=\norm{\widehat{P}_0^c \widehat{\vec{J}^c}+ \widehat{\vec{P}^c}\times\widehat{\vec{B}^c}}^2
	\end{equation}
	where the eigenvalue of $\widehat{\mathfrak{I}}_1$ is given by the energy $E$ of the Carroll particle and $\widehat{\mathfrak{I}}_2$ is the ultra-relativistic analogue of square of Pauli-Lubanski pseudo-vector.
	\subsubsection{Dispersion relation}
	Let us now shed some light on the ultra-relativistic limit of the dispersion relation $	E^2=
	\left(\overrightarrow{p}^2 c^2 + m^2 c^4\right)$.  Now putting $\vec{p} = \gamma m \vec{v}$ in the R.H.S, where 		$\gamma=\frac{1}{\sqrt{1-\frac{\vec{v}^2}{c^2}}}$ we can rewrite the relativistic dispersion relation as following:
	\begin{equation}
		\frac{E^2}{\vec{p}^2c^2}
		= 1 + \left(1-\frac{v^2}{c^2}\right)
		\frac{1}{\left(\frac{v^2}{c^2}\right)}
		\quad;\quad
		v := |\vec{v}|
	\end{equation}
	Now the ultra-relativistic regime is defined as $v\to c$ limit and as the velocity of a particle $v < c$, the limit $c\to 0$ must be accompanied by the limit $v\to 0$ simultaneously such that $\frac{v}{c} \,\to 1.$ With this limit the above dispersion relation in the $c\to 0$ limit reduces to
	\begin{equation}
		\left(\frac{E^2}{\vec{p}^2c^2}\right)
		\to 1\,\,
		\implies \, 
		E \to \left|\overrightarrow{p}\right| c\label{dispersion carroll}
	\end{equation}
	which is the so called ultra-relativistic or Carrollian dispersion relation.
	\section{$\kappa$- Minkowski space and deformed phase space algebra: A revisit}		  
	To accomplish our target, as spelled out in the introduction, we take a top-down approach in the remaining sections. We employ the Bopp map for $\kappa$-Minkowski space-time (refer to equation (3.36) of   \cite{biswajit2023symmetries}) albeit at the \textit{quantum level}. Similar to the approach outlined in Section 2, where we had derived Galilean and Carrollian representations of the symmetry generators, we will obtain the corresponding deformed generators of the $\kappa$-Minkowski space-time using the relativistic (finite $c$) Bopp map in terms of Galilean and Carrollian coordinates. Subsequently, by taking the appropriate limits as the speed of light $c\to\infty$ or $0$, we obtain the corresponding expressions for the limiting version of the Bopp map, bridging the gap between commutative and non-commutative coordinates. Through this process, we extract the components of the momentum space tetrad. Utilizing these components, we calculate the metric of the deformed subspace of the momentum space, enabling us to compute the deformed dispersion relations in the respective cases. \\ \\
	In this exploration, a noteworthy aspect emerges concerning the fate of non-commutative parameters $a^{\mu}$ in the respective scenarios. As we shall see that during the process of extracting the Bopp map for $\kappa$-Minkowski space in non-relativistic and ultra-relativistic limits, a rescaling of both the temporal ($a^0$) and spatial ($a^i$) components of the non-commutative parameter $a^{\mu}$, like some of the ISO(1,3) generators, becomes necessary for a viable limit ($c\to\infty$ and $c\to 0$ respectively) to exist. This forces us to redefine the non-commutative parameters before taking the respective limits. Intriguingly, this process naturally leads to the selection of a purely space-like or purely time-like non-commutative parameters for the Galilean and Carrollian cases, respectively giving rise to a fundamental length scale ($a^i$) and time scale ($a^0_c$) in the corresponding regimes.
	Presumably, they are of the Planck length and Planck time scale 
	respectively.
	\subsection{Non-relativistic limit of $\kappa$-Minkowski space-time : $\kappa$-Galilean space}
	Hitherto we were occupied with the construction of Galilean and Carrollian limits of commutative space-time. We now extend this study to NC $\kappa$-Minkowski space-time $\widehat{\mathcal{M}}$. For that, let us first start with the construction of $\kappa$-Galilean space-time and first try to write down the non-commutative coordinate algebra for Newton-Cartan space-time $\widehat{\mathcal{N}}$. To define the operator valued commutative Galilean coordinates, we elevate the Galilean coordinates $q_g^{\mu}$ \eqref{gcoordinates} to the level of operators as $q^{\mu}_g\,\,\to\,\,\widehat{q}^{\mu}_g$ satisfying $[\hat{q}^{\mu}_g,\hat{q}^{\nu}_g]=0$ appropriate for commutative space-time $\mathcal{N}$ and then again to the operator valued $\kappa$-Minkowski coordinates $\widehat{X}^{\mu}$ just like \eqref{gcoordinates} as 
	\begin{equation}
		\label{ncgc}
		\widehat{X}^0 \longrightarrow 
		\widehat{X}_g^0 
		= \lim_{c\to \infty} \widehat{T}(c)
		= \lim_{c\to \infty}\left(\frac{\widehat{X}^0}{c}\right);\qquad				\widehat{X}^i \longrightarrow 
		\widehat{X}_g^i = \widehat{X}^i
	\end{equation}
	To derive the coordinate brackets, we begin by expressing the $\kappa$-Minkowski coordinate brackets \eqref{def_minkow_eq1} in a non-covariant manner i.e. by splitting the temporal and spatial coordinates. To start with consider,
	\begin{align*}
		\numberthis
		\label{def_gal_eq1}
		\comm{\frac{\widehat{X}^0}{c}}{\widehat{X}^i}
		&= i\left(\frac{a^0}{c}\widehat{X}^i 
		- a^i\frac{\widehat{X}^0}{c}\right)
	\end{align*}
	Now if we take the limit $c\to\infty$  in both sides of \eqref{def_gal_eq1}, we get by holding $a^0$ finite, 
	\begin{equation}
		\comm{\widehat{X}^0_g}{\widehat{X}^i_g}=-ia^i\widehat{X}^0_g\label{coordinate1}
	\end{equation}
	On the other hand, the commutator in the spatial coordinate sector \eqref{ncgc} remains undeformed.
	\begin{equation}
		\comm{\widehat{X}^i_g}{\widehat{X}^j_g}=i\left(a^i\widehat{X}^j_g-a^j\widehat{X}^i_g\right)\label{coordinate2}
	\end{equation}
	These are the $\kappa$-Galilean algebra satisfied by the Galilean coordinates.\\
	In the next subsection we start with the Bopp map for $\kappa$-Minkowski space-time given in   \cite{biswajit2023symmetries} and apply the coordinate transformation \eqref{ncgc} to derive the corresponding Bopp map for $\kappa$-Galilean space-time and with this we can verify the above algebra. Eventually the Galilean Bopp map will enable us to derive the phase-space algebra, which, in turn, will be shown to impact single-particle dispersion relation in a non-trivial manner.
	\subsubsection{Bopp map for $\kappa$-Galilean coordinates}
	\noindent
	A non-canonical mapping (Bopp map) between the $\kappa$-Minkowski coordinates $X^{\mu}$ and commutative Minkowski coordinates $q^{\mu}$, at the classical level, was obtained in   \cite{biswajit2023symmetries} as
	\begin{equation}
		\label{33}
		X^{\alpha}
		= \left(E^{-1}(P)\right)^a\,_{\mu}q^{\mu} = \left(q^{\alpha}\Phi(P)-(a\cdot q)P^{\alpha}\right);
		\qquad 	
		\Phi(P)
		=\left(a^\mu P_\mu + 
		\sqrt{1+a^\mu a_\mu P^\nu P_\nu}\right)
	\end{equation}
	However, upon closer inspection, it becomes evident that there is no hurdle on the way for us to make a seamless transition 
	to the quantum operatorial level, establishing a direct connection 
	between the commutative and non-commutative $\kappa$-Minkowski 
	coordinate \emph{operators} satisfying the commutation relations \eqref{def_minkow_eq1}. In order to obtain a Bopp 
	map connecting the $\kappa$-Galilean coordinate $\hat{X}^{\mu}_g$ and 
	the commutative Galilean coordinates $\widehat{q}^{\mu}_g$, we can therefore employ the same mapping \eqref{33} directly at the quantum level, 
	eliminating the need for any consideration of operator ordering. It is evident that even at the classical level the vector fields $P_{\mu}$'s in \eqref{33} act on $\phi(q) \,\in \,C^{\infty}(\mathcal{M})$. So in order to make it act on the space of Schr\"{o}dinger fields $\psi(\vec{q}_g,q_g^0)\in C^\infty(\mathcal{N})$ it is important to transform them first unitarily. Correspondingly, these space-time operators $\widehat{X}^\alpha$ should also transform just like $\mathfrak{iso}$(1,3) generators adjointly as in \eqref{rev_m_gal_eq4d}\footnote{Note that $\hat{q}_i,\hat{P}_i$ will remain unaffected. Only $\hat{P}_0$ will transform to $\hat{\tilde{P}}_0=\frac{\hat{P}_0^g}{c}$ \eqref{q8} which is implemented in the next step.}. So the $\widehat{X}^\alpha$ \eqref{33} along with $\widehat{P}^\alpha$ needs to transform as.  
	\begin{equation}
		\label{34}
		\widehat{X}^\alpha\longrightarrow \widehat{\tilde{X}}^{\alpha} = e^{imc^2t}\widehat{X}^\alpha e^{-imc^2t} = \left(\tilde{q}^{\alpha}\Phi\left(\widehat{\tilde{P}}\right)-(a\cdot\tilde{q})\widehat{\tilde{P}}^{\alpha}\right); \myquad[2]
		\widehat{P}^\alpha\longrightarrow \widehat{\tilde{P}}^\alpha = e^{imc^2t}\widehat{P}^\alpha e^{-imc^2t}
	\end{equation}We begin by looking at  the spatial index, $\mu=i$, for which the map \eqref{33} reduces to the 
	following if written in terms of $\widehat{P}_0^g$ and $t(c)$ .
	\begin{align*}
		\numberthis
		\label{bopp1}
		\widehat{\tilde{X}}^i = q^i \left\{\left(a^0\widehat{\tilde{P}}_0 - \vec{a} \cdot \widehat{\overrightarrow{P}}\right)
		+\sqrt{1+\Big((a^0)^2 - (\vec{a})^2\Big)
			\left(\left(\widehat{\tilde{P}}_0\right)^2
			- \left(\widehat{\overrightarrow{P}}\right)^2
			\right)}\right\}-\big(c a^0 t - 
		\vec{a} \cdot \vec{q}\big)\widehat{P}^i
	\end{align*}
	Now in the limit $c\to \infty$ the term $\widehat{\tilde{P}_0}=\frac{1}{c}\left(c\widehat{\tilde{P}}_0\right)\,\to \frac{1}{c}\widehat{P}_0^g=0$ giving vanishing result and the term $\left(ca^0t\widehat{P}^i\right)$ diverges. So to make it converge we need to put $a^0=0$. So by identifying $\widehat{X}^i_g:=\widehat{\tilde{X}}^i$ in the limit $c\to\infty$ in \eqref{bopp1}, we get 
	\begin{align*}
		\numberthis
		\label{def_gal_eq7}
		\widehat{X}_g^i
		&= q_g^i \left(-\vec{a}\cdot \widehat{\overrightarrow{P}}_g
		+ \sqrt{1 + \vec{a}^2 \widehat{\overrightarrow{P}}_g^2}
		\right)
		+ (\vec{a}\cdot \vec{q}_g) \widehat{P}^i_g
	\end{align*}
	Now let us consider the temporal component $\mu=0$. In \eqref{34} we set $a^0=0$ right from the beginning and then implement the above mentioned unitary transformation where $\widehat{X}^0 \,\to\,\widehat{\tilde{X}}^0$ and $\widehat{P}_0\,\to\,\widehat{\tilde{P}}_0$ while $q^0,\widehat{P}_i$ are not transformed, to get 
	\begin{align*}
		\numberthis
		\label{bopp2}
		\widehat{\tilde{X}}^0 = q^0 \left\{-\vec{a} \cdot \widehat{\overrightarrow{P}}
		+\sqrt{1 - \vec{a}^2
			\left(\left(\widehat{\tilde{P}}_0\right)^2
			- \left(\widehat{\overrightarrow{P}}\right)^2
			\right)}\right\}+\big(
		\vec{a} \cdot \vec{q}\big)\widehat{\tilde{P}}^0
	\end{align*}
	In the limit $c\to\infty$, $\frac{\widehat{\tilde{X}}^0}{c}\,\,\to\,\,\widehat{X}_g^0$ and $\frac{q^0}{c}\,\,\to\,q^0_g$ and $\widehat{\tilde{P}}_0\,\to\,0$, leaving us with
	\begin{align*}
		\numberthis
		\label{eq24}
		\widehat{X}_g^0 
		&= q_g^0 \left(-\vec{a}\cdot \widehat{\overrightarrow{P}}_g
		+ \sqrt{1 + \left(\vec{a} \right)^2 \left(\widehat{\overrightarrow{P}}_g\right)^2 
		}
		\right)
	\end{align*}
	So \eqref{def_gal_eq7} and \eqref{eq24} furnishes us with the relation between non-commutative and commutative Galilean coordinates. We can also recheck at this stage that using these two expressions and the canonical brackets between $q$'s and $P$'s the $\kappa$-Galilean coordinate algebras \eqref{coordinate1},\eqref{coordinate2} are indeed reproduced. To summarize the lessons learnt from this exercise is that for the Galilean limit of the $\kappa$-Minkowski space-time to exist, it is mandatory to consider a \textit{purely spacelike}  non-commutative parameter $a^{\mu}$  i.e. has no temporal component ($a^0=0$). From now onwards we shall suppress the super/subscripts $g$
	and denote the Galilean generators and coordinate operators simply as 
	$\widehat{J}_{ij},\widehat{B}_i,\widehat{P}_\nu,\widehat{X}^\mu$.
	\\ \\
	Now with this Bopp map \eqref{def_gal_eq7}, 
	\eqref{eq24} we can calculate the phase-space algebra 
	$\comm{\widehat{P}_\nu }{\widehat{X}^\mu}$, in a straight forward manner to get
	\begin{subequations}
		\label{phasespaceg}
		\begin{align*}
			\numberthis
			\comm{\widehat{P}_0}{\widehat{X}^0}
			&= -i\phi\left(\vec{a},\widehat{\overrightarrow{P}}\right);\quad\quad 	\phi\left(\vec{a},\widehat{\overrightarrow{P}}\right) = 
			-\vec{a}\cdot\widehat{\overrightarrow{P}}
			+\sqrt{1 + (\vec{a})^2\left(\widehat{\overrightarrow{P}}\right)^2}
			\\
			\numberthis
			\comm{\widehat{P}_0}{\widehat{X}^k}
			&= 0;\qquad 
			\comm{\widehat{P}_i}{\widehat{X}^0}
			= 0
			\\
			\numberthis
			\comm{\widehat{P}_i}{\widehat{X}^k}
			&= 
			-i\tensor{\delta}{_i^k}
			\phi\left(\vec{a},\widehat{\overrightarrow{P}}\right)
			+ ia_i\widehat{P}^k
		\end{align*}
	\end{subequations}
	It is crucial to note that in the non-relativistic case, the RHS of these expressions involve only the spatial components of the momentum operator. However, in the ultra-relativistic scenario, we shall uncover a completely contrasting feature. 
	\\
	We can also derive the brackets between $\widehat{B}_i$'s and $\widehat{X}^{\mu}$'s using the Bopp map and the commutation relation between the  commutative Galilean coordinates and the boost generators.
	\begin{subequations}
		\label{eq5}
		\begin{align}
			\comm{\widehat{J}_{ij}}{\widehat{X}^0}
			&= i \phi^{-1}\left(a_i \widehat{P}_j - a_j \widehat{P}_j\right)\widehat{X}^0
			\\
			\comm{\widehat{J}_{ij}}{\widehat{X}^k}
			&= i \left(\tensor{\delta}{_i^k}\widehat{X}_j - \tensor{\delta}{_j^k}\widehat{X}_i\right) - i\left(a_i\tensor{\widehat{J}}{_j^k} - a_j \tensor{\widehat{J}}{_i^k}\right) 
			%+ \left(a_j\tensor{\delta}{_i^k} - a_i \tensor{\delta}{_j^k} \right)
			\\ 
			\comm{\widehat{B}_i}{\widehat{X}^0}
			&= i m \widehat{X}^0
			\phi^{-1}
			\left(
			-a_i 
			+
			\vec{a}^2 \left(\phi + \vec{a}\cdot\widehat{\overrightarrow{P}}\right)^{-1}\widehat{P}_i
			\right)
			%%%%%%%%%%%%%%
			%%%%%%%%%%%%%%
			\cmnt{\comm{\widehat{B}_i}{\widehat{X}^0}
				&= i m \widehat{X}^0
				\left(-\vec{a}\cdot\widehat{\overrightarrow{P}}
				+ \sqrt{1+\vec{a}^2  
					\widehat{\overrightarrow{P}}^2}\right)^{-1}
				\left(
				-a_i 
				+
				\vec{a}^2 \left(1+\vec{a}^2\widehat{\overrightarrow{P}}^2
				\right)^{\sfrac{-1}{2}}
				\widehat{P}_i
				\right)}
			%\\ \nonumber
			%%%%%%%%%%%%%%
			%%%%%%%%%%%%%%
			\\ 
			\begin{split}
				\comm{\widehat{B}_i}{\widehat{X}^j}
				&= i\tensor{\delta}{_i^j}
				\widehat{X}^0 + ia_i\widehat{B}^j
				- \frac{im}{\phi\left(\phi+\vec{a}\cdot \widehat{\overrightarrow{P}}\right)}\Bigg\{ -\vec{a}^2\widehat{P}_i\widehat{X}^j 
				+ \left(\tensor{\delta}{_i^j}\phi + \vec{a}^2\frac{\widehat{P}_i\widehat{P}^j }{\phi + \vec{a}\cdot\widehat{\overrightarrow{P}}}\right)\left(\vec{a}\cdot\widehat{\overrightarrow{X}}\right)
				\Bigg\}
			\end{split}
		\end{align}
	\end{subequations}	
	It is evident that the algebra of coordinates with the rotation and boost generators are highly deformed and in fact they are enveloping algebra valued.
	As a consistency check, it can also be verified that the Jacobi identities among any triplets of generators of this extended universal enveloping algebra formed by Galilean generators and the space-time coordinate operators($\widehat{J}_{ij},\widehat{B}_{k},\widehat{P}_\mu,\widehat{X}^\mu$) are satisfied.

	\subsubsection{Momentum space of the $\kappa$-Galilean space and deformed dispersion relation}
	From the previous section we can see that the translation of the $\kappa$-Galilean coordinates depend on the spatial components of momentum, indicating that the corresponding momentum space might be of non-trivial nature. In this section we shall first determine the metric of the momentum space following the approach of   \cite{biswajit2023symmetries} and then derive a deformed dispersion relation for $\kappa$-Galilean particle. As the analysis has to be a classical one, we first demote the operator valued coordinates to the level of classical i.e. commuting variables and then demote the required commutator brackets to the level of classical Dirac brackets. The corresponding brackets will correspond to the deformed symplectic structure of a classical particle living in a $\kappa$-Galilean space-time\footnotemark. To that end, we begin by demoting the commutator brackets to the level of Dirac brackets as following:
	\footnote{Again as in   \cite{biswajit2023symmetries}, we can write down a Lagrangian linear in ``velocities" reproducing the deformed symplectic structures, associated with Dirac brackets stemming from the second class constraints. Alternatively, these Dirac brackets can also be determined directly using the formalism developed in \cite{faddeev-jackiw}. However, we do not make any effort to construct this Lagrangian here as our primary goal is to study the fate of dispersion relation in these limits from the first principle.}
	\begin{align*}
		\numberthis
		\label{db}
		\comm{f\left(q,\widehat{P}\right)}{g\left(q,\widehat{P}\right)}
		\longrightarrow \db{f(q,P)}{g(q,P)} =
		\displaystyle{\lim_{\hbar\to 0}
			\frac{\comm{f\left(q,\widehat{P}\right)}{g\left(q,\widehat{P}\right)}}
			{i\hbar}}
	\end{align*}
	So after reinstating the $\hbar$ in the phase-space algebra \eqref{phasespaceg} in suitable places and then using \eqref{db} in \eqref{coordinate1},\eqref{coordinate2} and \eqref{phasespaceg} we arrive at the following Dirac brackets
	\begin{subequations}
		\begin{align}
			\db{X^0}{X^j}
			&= -\mathfrak{a}^j X^0;\qquad
			\db{X^i}{X^j}
			= \left(
			\mathfrak{a}^i X^j 
			- \mathfrak{a}^j X^i
			\right)
		\end{align}
	\end{subequations}
	\begin{subequations}
		\label{eq27}
		\begin{align*}
			\numberthis
			\db{P_0}{X^0}
			&= \left(\vec{\mathfrak{a}}\cdot\vec{P}
			-
			\sqrt{1 + 
				\vec{\mathfrak{a}}^2 \vec{P}^2}\right)
			\\
			\numberthis
			\db{P_0}{X^k}
			&= 0;\qquad
			\db{P_i}{X^0}
			= 0
			\\
			\numberthis
			\db{P_i}{X^k}
			&= 
			\tensor{\delta}{_i^k}
			\left(
			\vec{\mathfrak{a}}\cdot\vec{P}
			-
			\sqrt{1 + \vec{\mathfrak{a}}^2\vec{P}^2}
			\right)
			+ \mathfrak{a}_i P^k
		\end{align*}
	\end{subequations}
	where  $\vec{\mathfrak{a}}
	=\displaystyle{\lim_{\hbar\to 0}\left(\frac{\vec{a}}{\hbar}\right)}$. Notice that as the limit $\hbar \to 0$ is approached, it becomes necessary to simultaneously consider $\vec{a} \to 0$ in such a way that the ratio $\frac{\vec{a}}{\hbar}$ remains fixed. This condition introduces a novel scale into the theory: while the Planck length diminishes ($a^i \sim \sqrt{\hbar G} \to 0$ (in $c=1$ unit)), a distinct mass scale emerges in the form of $\mathfrak{a}\sim\sqrt{\frac{G}{\hbar}}$ (presumably corresponding to the inverse of the Planck mass scale $m_P$).
	\\ \\
	Now from \eqref{eq27}, we can just read off the components of inverse of the momentum space tetrad $E^{-1}(\vec{P})$ as was introduced in \eqref{tetrad} to get
	\begin{equation}
		\tensor{E^{-1}\left(\vec{P}\right)}{^0_0}
		= 
		\left(-\vec{\mathfrak{a}} \cdot \vec{P}
		+
		\sqrt{1 +
			\vec{\mathfrak{a}}^2
			\vec{P}^2}
		\right);\qquad
		\tensor{E^{-1}\left(\vec{P}
			\right)}{^i_j}
		=  \tensor{\delta}{^i_j}
		\left(
		-\vec{\mathfrak{a}} \cdot \vec{P}
		+
		\sqrt{1 +
			\vec{\mathfrak{a}}^2
			\vec{P}^2}
		\right)
		- \mathfrak{a}_j P^i\label{3.16b}
	\end{equation} 
	Now just like in   \cite{biswajit2023symmetries}, we can use this momentum space tetrad to find the information abut metric of the momentum space of a non-relativistic free particle. As per the protocol described in   \cite{biswajit2023symmetries} we first work out the momentum space metric of the Galilean algebra \eqref{galalg} from its Casimir relation. In the context of the massless case (Galilean algebra), it is simply given by $(\widehat{\vec{P}}\,^g)^2$ \eqref{a1}, whereas for the massive case (Bargmann algebra), it is formally given by $\left(P_0-\frac{\vec{P}^2}{2m}\right)$, albeit with divergent $P_0$. By Schur's lemma this should be proportional to identity and using \eqref{A7} the proportionality constant can be identified with the constant divergent quantity $mc^2$  (in the case $c \to\infty$ limit). But their difference $(P_0-mc^2)$ is finite in this limit, as occurs in the non-relativistic dispersion relation \eqref{A7}. In either cases, since the quadratic form of the Casimir is not there in the entire energy-momentum space, which has a quadratic part involving only the 3-momenta $\vec{P}$, we can infer that it is only the 3D momentum space which inherits the flat Euclidean geometry with the associated metric is simply $\eta^p_{ab}=-\delta_{ab}$,
	whereas, $\eta_{00}^p$ remains undefined.
	This means that although the corresponding momentum space is topologically four dimensional, the metric property can be seen to be inherited only by the 3-dimensional subspace of  spatial momentum.\\ \\
	Now using \eqref{3.16b} the curved 3D momentum space metric can be obtained as,  
	\begin{align*}
		\numberthis
		\label{g0}
		g^p_{ij} 
		= \eta^p_{ab} \tensor{(E^{-1})}{^a_i}
		\tensor{(E^{-1})}{^b_j}
	\end{align*}
	where, $\eta^p_{ab}$ is the flat momentum space metric 
	corresponding to local tangent plane of the curved momentum space \footnote{Here, we have used early Latin indices $a,b,..$ etc. to indicate the orthonormal basis for the 3D local tangent space and the middle Latin indices like $i,j,...$ to indicate world indices of 3D momentum space.}.	Thus, the components of the momentum space metric of the $\kappa$-Galilean particle is given by
	\begin{align}
		\label{g1a}
		g^p_{ij}(\vec{P})
		&= \phi
		\left(
		\phi\delta_{ij}
		-\mathfrak{a}_i P_j
		- \mathfrak{a}_j P_i
		\right)
		+ \mathfrak{a}_i \mathfrak{a}_j \vec{P}^2
	\end{align}
	where, we have dropped the irrelevant minus sign and $\phi$ stands for the following object:
	\begin{align}
		\label{g2}
		\phi(\vec{P};\vec{\mathfrak{a}})
		=	-\vec{\mathfrak{a}}\cdot\vec{P}
		+
		\sqrt{1 + \vec{\mathfrak{a}}\,^2 \vec{P}^2}
	\end{align}
	Now this metric can be used to derive the geodesic distance between a arbitrary point $P\,\in\,\mathcal{P}$-the curved 3D momentum space and a suitably chosen origin $O$ in $\mathcal{P}$ (analogue of sec-3.1 of   \cite{biswajit2023symmetries} but in the 3D spatial subspace). However this should not be identified with the pull-back of the metric $g_{\mu\nu}$ of   \cite{biswajit2023symmetries} (see equation (3.22) in   \cite{biswajit2023symmetries}) in 3-dimension as the metric becomes degenerate in the temporal sector and due to this reason the time-like geodesic in the 4-dimensional curved momentum space may not go over to the geodesic in the 3-dimensional momentum space defined by the metric \eqref{g1a}. Indeed in the latter case the geodesic is completely space-like. The squared distance in Newton-Cartan flat 3D space, which is $C=\delta_{ij}P^{i}P^{j}=\vec{P}^2$, generalises   \cite{carmona,relancio} in curved momentum space to
	\begin{equation}
		C=[D(0,P)]^2=\left(\int_0^P\,\,\sqrt{g^p_{ij}(p)dp^{i}dp^{j}}\right)^2=\left(\int_0^{\tau}\,\,d\tau '\, \sqrt{g^p_{ij}(p)\dot{p}^{i}\dot{p}^{j}}\right)^2\,\,;\qquad \dot{p}^{i}=\frac{dp^{i}}{d\tau '}\label{g3}
	\end{equation}
	However this object is not invariant under an arbitrary diffeomorphism of $\mathcal{P}$ because of the presence of $\vec{a}$, which is a triplet of just 3 numbers and takes same values in all frames; it does not transform as a 3-vector even under a simple SO(3) rotation.
	So to circumvent this problem we elevate $a^i$'s to a 3-vector under diffeomorphism in $\mathcal{P}$ as an interim measure. With this the metric $g_{ij}$ is promoted to a proper tensor, rendering $C$ to be a scalar. Now treating $\mathcal{P}$ to be like a genuine Riemannian manifold, we make use of the following differential equation   \cite{thanu} satisfied by $C(\vec{p}^2)$ :
	\begin{equation}
		\partial^{\mu}C(\vec{p}\,^2)\,g^p_{\mu\nu}(\vec{p})\,\partial^{\nu}C(\vec{p}\,^2)=4C(\vec{p}\,^2)\label{g4}
	\end{equation}
	Note that here we have considered that $C$ is a function of $\vec{P}^2$, reflecting the fact that two different points $P_1$ and $P_2$ with coordinates $\vec{P}_1$ and $\vec{P}_2$, satisfying the same `shell' condition $\vec{P}_1^2=\vec{P}_2^2=K^2$ will have the same geodesic distance from the chosen origin $O$. It ensures that the deformed $C$ will also be a part of Casimir of the Galilean algebra and will be invariant under Galilean transformations.  The differential equation \eqref{g4} then reduces to 
	\begin{equation}
		\int_0^{P^2}\frac{dC(\Vec{p}\,^2)}{\sqrt{C(\Vec{p}\,^2)}}=\int_0^{P^2}\frac{d(\Vec{p}\,^2)}{\sqrt{(1+\vec{\mathfrak{a}}^2\Vec{p}\,^2)\Vec{p}\,^2}}\label{g5}
	\end{equation}
	where we have used the following identity
	\begin{equation}
		g^p_{ij}p^{i}p^{j}=(1+\vec{\mathfrak{a}}^2\vec{p}\,^2)\vec{p}\,^2
	\end{equation}
	Now solving the integral equation in \eqref{g5}, we arrive at
	\begin{equation}
		C(\vec{P}^2)=D^2=\frac{1}{\mathfrak{a}^2}\left[\sinh^{-1}\Big(\mathfrak{a}|\vec{P}|\Big)\right]^2;\quad \mathfrak{a}:=|\vec{\mathfrak{a}}|\label{g6}
	\end{equation}
	It can be verified easily at this stage that as $\mathfrak{a}\to 0,\,\,C(\vec{P}^2) \to \vec{P}^2$.\\
	We therefore just need to replace $\vec{P}^2\,\, \rightarrow\,\, C(\vec{P}^2)$, along with a deformed energy $E(\mathfrak{a})$ (with $E(\mathfrak{a}\to 0)\to E_p$) in the undeformed dispersion relation $\left(E_p - \frac{\vec{P}\,^2}{2m}\right)=0$ \eqref{A7} to get the corresponding deformed one.  The generalized Casimir and dispersion relation for the curved momentum space $\mathcal{P}$ is then given by 
	\begin{equation}
		E(\mathfrak{a})-\frac{1}{2m\mathfrak{a}^2}\left[\sinh^{-1}(\mathfrak{a}|\vec{P}|)\right]^2=0
		\label{a2}
	\end{equation}
	where one can take this equation \eqref{a2} as the definition of $E(\mathfrak{a})$ itself. 		
	Note further that we can interpret $C(\vec{P}^2)$ as $\vec{\Pi}^2$, in  the spirit of   \cite{biswajit2023symmetries}, where $\vec{\Pi}$ is a three momentum vector representing the Riemann  normal coordinates defined at the tangent plane $T_O(\mathcal{P})$ anchored at the origin $O$ of $\mathcal{P}$. With this the deformed dispersion relation simply becomes
	\begin{equation}
		\label{3.27}
		\left(\Pi_0+\frac{\vec{\Pi}^2}{2m}\right)=0 
		\quad ; \quad\Pi_0 := -E(\mathfrak{a})
	\end{equation}
	where unlike in the relativistic case   \cite{biswajit2023symmetries} the mass $m$ undergoes no renormalization. On the other hand $\Pi_0$ is a suitable deformation of $P_0$ so that the left hand side of \eqref{3.27} can be thought of as a new Casimir under the deformed Galilean boost  $B_i(\mathfrak{a})=-\left(
	q^0\Pi_i + mq_i
	\right)$, obtained from \eqref{rev_m_gal_eq4a} by replacing $P_i \rightarrow \Pi_i$ and operates at $T_O(\mathcal{P})$, where it can be shown to commute with $\left(E(\mathfrak{a})-\frac{\vec{\Pi}^2}{2m}\right)$ which is nothing but the counter part of \eqref{gcasimir}. In the commutative limit $E(\mathfrak{a}) \,\to \, E$ and $\vec{\Pi}\,\to\,\vec{P}$ and,
	the fact that $\Pi(\vec{P})$ is a non-linear function of $\vec{P}$ is indicative of non-linear composition (addition) rule of linear momenta. This should be clear in a multi-particle system, where commutativity and associativity may not hold anymore, giving rise to torsion and curvature in the curved 3d momentum space   \cite{glickman_curved_mom_space,glickman_intro_to_dsr}.
	\\\\	
	Again note in this context that a naive execution of the limit $c\to \infty$, for the space-like non-commutative parameter $\mathfrak{a}^2 < 0$ in (3.36) of   \cite{biswajit2023symmetries} is not sensible because of the presence of $\sin^{-1}(\mathfrak{a}mc)$ (where we have reinstated $c$ within the parenthesis from dimensional consideration). This is not unexpected after all as the momentum space metric in the parent 4-dimensional case becomes singular in this limiting situation and cannot expect the time-like geodesic (3+1)D momentum space to go over to a space-like one in $c\to\infty$ limit. 
	\subsection{Ultra-relativistic limit of $\kappa$-Minkowski space-time : $\kappa$-Carroll space-time}
	In this section, we will take the \emph{ultra-relativistic limit} of the space-time commutation algebra in \eqref{def_minkow_eq1}. For obtaining the operator valued Carroll coordinates satisfying a non-commutative Lie algebra from the Minkowski coordinates we follow \eqref{carroll}. The non-commutative Carroll coordinates $\widehat{X}^{\mu}_c$ can be obtained as 
	\begin{align*}
		\numberthis
		\label{def_carr_m_eq3}
		\widehat{X}^0 &\longrightarrow 
		\widehat{X}_c^0 
		= \lim_{c\to 0} \widehat{T}(c)
		= \lim_{c\to 0}\left(\frac{\widehat{X}^0}{c}\right)
		;\myquad[2]
		\widehat{X}^i \longrightarrow 
		\widehat{X}_c^i = \widehat{X}^i
	\end{align*}
	Now to find the space-time commutator algebra in the Carrollian limit we first start with the following commutator from \eqref{def_minkow_eq1}:
	\begin{align*}
		\numberthis
		\label{def_carr_m_eq4}
		\lim_{c\to 0}
		\comm{\frac{\widehat{X}^0}{c}}{\widehat{X}^j}
		&= i
		\lim_{c\to 0}\left(\left(\frac{a^0}{c}\right)\widehat{X}^j - a^j \frac{\widehat{X}^0}{c}\right)
	\end{align*}
	To retain the contribution from the otherwise diverging first term, it becomes imperative to ensure that $a^0 \to 0$ simultaneously so that the ratio $\frac{a^0}{c}$ is held fixed to a specific constant value say $a^0_c$. In the ultra-relativistic regime, although the temporal non-commutative parameter $a^0$ approaches zero, a new time scale  emerges through $a^0_c$ and, is presumably of likely the order of \textit{realistic}  the Planck time $t_p= \sqrt{\frac{\hbar G}{c^5}}\sim 10^{-44}$ secs\footnotemark. 
	%%%%
	\footnotetext{By the adjective ``realistic", we mean that all the constants like $\hbar,\, G,$
		and $c$ take their realistic values in the definition of Planck time $t_p$ here and the \textit{formal} mathematical limits like $\hbar, c \to 0$ are not taken.}
	%%%%
	Thus the 4 deformation parameter(4-scalars) also get 
	transformed as, 
	\begin{align}
		\label{def_carr_m_eq7}
		a^0 &\longrightarrow a_c^0
		= \lim_{c\to 0} a^t(c)
		=\lim_{c\to 0}\left(\frac{a^0}{c}\right)	\quad\textrm{where}\qquad
		a^t(c) = 
		\frac{a^0}{c}
		\\
		a^i &\longrightarrow a_c^i = a^i
	\end{align}
	Then, \eqref{def_carr_m_eq4} in the limit $c\to 0$ eventually takes the form,
	\begin{subequations}
		\begin{align*}
			\numberthis
			\label{def_carr_m_eq5}
			\comm{\widehat{X}_c^0}{\widehat{X}_c^j}
			&= i\left(a_c^0 \widehat{X}_c^j 
			- a_c^j \widehat{X}_c^0\right),
			\intertext{Also, while the spatial Carrollian coordinate
				operators retains their undeformed form \eqref{def_minkow_eq1}}
			\numberthis
			\label{def_carr_m_eq6}
			\comm{\widehat{X}_c^i}{\widehat{X}_c^j}
			&= i\left(a_c^i \widehat{X}_c^j 
			- a_c^j \widehat{X}_c^i\right)
		\end{align*}
	\end{subequations}

	\subsubsection{Bopp map for $\kappa$-Carrollian coordinates}
	%Code: def_carr_alg
	In this subsection we shall now derive the mapping between the commutative and non-commutative Carroll coordinates and for that we follow the same procedure as shown in the Galilean case, discussed in the previous section. To that end, we consider the particular equation \eqref{33} and first consider its $\mu=0$ component to get the following equation
	\begin{align*}
		\numberthis
		\frac{\widehat{X}^0}{c} =
		\frac{q^0}{c}\left\{\left(a^t\widehat{P}_t - \vec{a} \cdot \widehat{\overrightarrow{P}} \right)
		+\sqrt{
			1+\big(c^2(a^t)^2 - \vec{a}^2\big)
			\left(\frac{\widehat{P}_t^2}{c^2}
			- \widehat{\overrightarrow{P}}^2
			\right)
		}\right\}-\big(c^2 a^t t - \vec{a}\cdot\vec{q}\big)
		\frac{\widehat{P}^0}{c}
	\end{align*}			
	Now in the limit $c\to 0,\quad t(c)\,\to\,q^0_c,\quad \frac{\widehat{X}^0}{c}\,\to\,\widehat{X}^0_c,\quad  c \widehat{P}_0 \to \widehat{P}_0^c, $ and  $a^t\,\to\,a^0_c,\,\, \Vec{a} \to \Vec{a}_c $ (see \eqref{carroll}, \eqref{def_carr_m_eq3},\eqref{rev_carr_alg_eq5a},\eqref{def_carr_m_eq7}) so that we can rewrite the above expression as ,
	\begin{align}
		\label{q1}
		\widehat{X}_c^0
		&= q^0_c \left\{\left(
		a_c^0 \widehat{P}_0^c - \vec{a}_c \cdot \widehat{\overrightarrow{P}_c}\right)
		+ \lim_{c\to 0}
		\sqrt{1+\big(c^2(a^t)^2 - (\vec{a}_c)^2\big)
			\left(\frac{\left(\widehat{P}_t\right)^2}
			{c^2}
			- \left(\widehat{\overrightarrow{P}_c}\right)^2\right)}\right\}
		- \lim_{c\to 0}\big(c^2 a^t t - \vec{a}_c \cdot \vec{q}_c\big)
		\frac{\widehat{P}_t}{c^2}
	\end{align}
	Now by inspection, the terms 
	$\left[\displaystyle{
		\left(\frac{\vec{a}_c^2}
		{c^2}\right)
		\left(\widehat{P}_t\right)^2
	}\right]$ inside the square root and the term
	$\left[\displaystyle{
		\left(\frac{a_c^j}
		{c^2}\right)
		q_c^j\widehat{P}_t
	}\right]$ are clearly divergent as 
	$c\to 0$. So to ensure its  convergence we \emph{must
		choose purely time-like} ($a^\mu$) so that only temporal component $a^0_c$ survives \eqref{def_carr_m_eq7} and the spatial components  $a^i_c$ vanishes: 
	\begin{align*}
		\numberthis
		\label{def_carr_alg_eq3}
		a_c^0 &\neq 0,
		\qquad
		\vec{a}_c=\vec{a}=0
	\end{align*}
	With this \eqref{q1} simplifies considerably to take the following form
	\begin{align*}
		\numberthis
		\label{q2}
		\widehat{X}_c^0
		&= \hspace{1mm}
		q_c^0\sqrt{1+(a_c^0)^2\left(\widehat{P}_0^c\right)^2}
	\end{align*}
	On the other hand, for the spatial components $\mu=i$ of \eqref{33}, if we employ the condition \eqref{def_carr_alg_eq3}, we then get the following equation
	\begin{align*}
		\numberthis
		\label{def_carr_alg_eq5}	
		\widehat{X}_c^i=
		q_c^i\left( a_c^0 \widehat{P}_0^c 
		+ \sqrt{1 + (a_c^0)^2\left(\widehat{P}_0^c\right)^2}
		\right)
	\end{align*}
	Equations, \eqref{q2} and \eqref{def_carr_alg_eq5}
	are the \emph{Bopp transformations} $\left(
	q_c^\mu \longleftrightarrow \widehat{X}_c^\mu\right)$ 
	in the \emph{ultra-relativistic} regime,
	i.e. the mapping between the commutative and the $\kappa$-Carrollian coordinates.
	\\ \\
	Now, applying \eqref{q2},\eqref{def_carr_alg_eq5}, we can recheck the following
	\begin{align*}
		\numberthis
		\label{def_carr_alg_eq6}
		\comm{\widehat{X}_c^0}{\widehat{X}_c^i}
		&= i a_c^0\widehat{X}_c^i \, ; \qquad
		\comm{\widehat{X}_c^i}{\widehat{X}_c^j}
		= \quad \mathbf{0}
	\end{align*}
	which matches with \eqref{def_carr_m_eq5},\eqref{def_carr_m_eq6} if we choose purely time-like $a^{\mu}_c$ \eqref{def_carr_alg_eq3}.	So it is crucial to note that,
	in the \emph{ultra-relativistic regime} of any physical system,
	i.e. for $c\to 0$, it is \emph{necessary} to have vanishing
	\emph{space-space non-commutativity} but 
	\emph{space-time} \emph{can be retained to be non-commuting}. In other words, it is found that \(a^i = 0\) and \(a^0 \to 0\) must hold, leading to a finite quantity \(a_c^0\).  
	\\ \\
	After obtaining  the \emph{invertible Bopp transformations} \eqref{q2},\eqref{def_carr_alg_eq5} within the realm of \emph{ultra-relativistic regime} our focus now shifts towards acquiring the \emph{deformed phase-space algebra}, paving the way for our journey into the intricacies of curved momentum space. Using the canonical commutators between the commutative coordinate $q^{\mu}_c$ and the momentum components we can easily derive the algebra for which the non-vanishing commutator are given as\footnote{From here onwards we shall drop the subscript or superscript `c' to designate Carrollian coordinate and simply put $a^0_c=a$ for simplicity.}
	\begin{align*}
		\numberthis
		\label{deformedcarroll}
		\comm{\widehat{P}_0}{\widehat{X}^0}
		= -i  \sqrt{1+a^2\left(\widehat{P}_0\right)^2}
		;
		\myquad[3]
		\comm{\widehat{P}_i}{\widehat{X}^k}
		= -i \tensor{\delta}{_i^k}
		\left(
		a\widehat{P}_0 + \sqrt{1+a^2\left(\widehat{P}_0\right)^2}
		\right)	
	\end{align*}
	which clearly involves only $\widehat{P}_0$, in the right hand sides of both the expressions in contrast to the Galilean case where it only involved the spatial components of momentum. \\
	As an aside to give a complete picture of the deformed transformation of the $\kappa$-Carrollian coordinates we can also calculate the commutators between the deformed Carrollian generators and the coordinate operators as 
	\begin{subequations}
		\label{eq13}
		\begin{align*} 
			\numberthis
			\comm{\widehat{J}_{ij}}{\widehat{X}^0} &=
			0
			&\comm{\widehat{J}_{ij}}{\widehat{X}^k}
			&= -i\left( \tensor{\delta}{_j^k}\widehat{X}_i
			-\tensor{\delta}{_i^k}\widehat{X}_j
			\right)
			\\ 
			\numberthis
			\comm{\widehat{B}_{i}}{\widehat{X}^0}
			&= -i  \left(
			\mathds{1} - a\left[ a\widehat{P}_0 
			+ \sqrt{1+a^2\left(\widehat{P}_0\right)^2}\right]^{-1}
			\hspace{1mm} \widehat{P}_0
			\right)\widehat{X}_i \, ,
			&\comm{\widehat{B}_{i}}{\widehat{X}^k}
			&= 0
		\end{align*}
	\end{subequations}
	The above equation shows that although the space-time rotation remain unaltered, however their transformation under deformed Carrollian boost are not Lie algebra valued and in fact valued in Universal enveloping algebra and depend on the zeroth component of momentum. And as a consistency check, it can also be verified that the Jacobi identities among any triplets of generators of this extended universal enveloping algebra($\widehat{J}_{\mu\nu},\widehat{P}_\mu,\widehat{X}^\mu$) formed by Carrollian generators and the space-time coordinate operators are satisfied.
	\cmnt{$= -i q_i 
		\sqrt{1+a^2\left(\widehat{P}_0\right)^2}$}
	\subsubsection{Construction of $\kappa$-Carrollian momentum space and deformed dispersion relation}
	Again in order to construct the momentum space metric we need to first demote the quantum operators to the level of classical variables and the commutators to the level of Dirac brackets(as done in \eqref{db}), where the non-vanishing brackets take the following form.					
	\begin{align}
		\label{q3}
		\db{X^0}{X^j}
		&= i\mathfrak{\mathfrak{a}}X^j,
		&\db{P_0}{X^0}
		&= -i  \sqrt{1+\mathfrak{a}^2\left(P_0\right)^2},
		&\db{P_i}{X^k}
		&= -i \tensor{\delta}{_i^k}
		\left(
		\mathfrak{a}P_0 + \sqrt{1+\mathfrak{a}^2\left(P_0\right)^2}
		\right)	
	\end{align}
	where all the Carrollian momentum components are defined to be conjugate momenta to the commutative Carrollian position components($q_c^0,q_c^i)$ and therefore defined to be in covariant nature. And, using \eqref{db} we have used $\mathfrak{a}=\displaystyle{\lim_{\hbar \to 0}}\,\, \left(\frac{a}{\hbar}\right)$. To interpret the aforementioned limit, it is necessary to consider the case where $a$ approaches zero. As we make the transition into the classical regime, the Planck time $a=a^0_c$  becomes infinitesimal, indicating the vanishing of this fundamental time scale. However, in this limit, a novel inverse energy scale denoted as $\mathfrak{a}$ comes into prominence.\\\\
	Now using \eqref{tetrad}, we can write down the components of the inverse momentum space tetrad from the phase-space algebra in \eqref{q3} as
	\begin{subequations}
		\label{tetradcarroll_1}
		\begin{align}
			\tensor{\left(E^{-1}(P_0)\right)}{^0_0}
			= \sqrt{1 + \mathfrak{a}^2(P_0)^2}
			\, \,
			,
			\qquad
			\tensor{\left(E^{-1}(P_0)\right)}{^i_j}
			= \tensor{\delta}{^i_j}
			\left(
			\mathfrak{a}P_0 + \sqrt{1 + \mathfrak{a}^2(P_0)^2}
			\right)
		\end{align} 
	\end{subequations}
	We can simply invert $E^{-1}(P_0)$ to get
	\begin{equation}
		\label{tetradcarroll_2}
		\tensor{\left(E(P_0)\right)}{_0^0}
		= \frac{1}{\sqrt{1 + \mathfrak{a}^2(P_0)^2}};\qquad					\tensor{\left(E(P_0)\right)}{_i^j}
		= \tensor{\delta}{_i^j}
		\frac{1}{\left(
			\mathfrak{a}P_0 + \sqrt{1 + \mathfrak{a}^2(P_0)^2}
			\right)}
	\end{equation}
	where we can replace $P_0^2 \to E^2$, with $E$ being the energy as $P_0^2$ is a Casimir of the algebra \eqref{rev_carr_alg_eq2}. Here, in contrast to the Galilean case, we will have to make use of the tetrad instead of inverse of the tetrads for reasons discussed in the following paragraph. Now to delve into the deformed momentum space of the non commutative Carrollian particle, we need to first find out the flat momentum space metric of a Carrollian particle. To that end we rewrite the eigen value equation of the Casimir operator as  
	\begin{equation}
		\label{carrcasimir}
		(P_0)^2=\eta_p^{\mu\nu}P_{\mu}P_{\nu}=E^2
	\end{equation}
	%	In \eqref{eq14}, it was earlier indicated that the dispersion relation for a Carrollian particle is given by $E=|\vec{P}c|$. The right hand side of the relation is held finite if we take $c\to 0$ and $|\vec{P}| \to \infty$ simultaneously. And indeed $E^2$ is a Casimir of the Carroll algebra given in \eqref{rev_carr_alg_eq2}. So we can assign an eigen value to the Casimir operator as $E^2=K^2$ from which we can extract out the metric of the corresponding momentum space with respect to which the inner-product is taken.
	where $\eta^{\mu\nu}_p$ represents the metric in the form of contravariant tensor of rank (2,0) of the Carrollian momentum space only. This is in contrast to the space-time metric of the Carroll space-time which was of rank (0,2) with $\cg_{00}=0$\, and \, $\cg_{ij}= -\delta_{ij}$ \eqref{C1}. From this, it becomes apparent that we can readily deduce $\eta_p^{00}=1$ and all other metric components $\eta_p^{\rho\sigma}\, (\rho,\sigma\neq 0)$ to be zero. It is important to note that, since the time component of the Carrollian momentum $P_0$ has no contravariant counterpart, so in order to contract the covariant momentum components with momentum space metric in \eqref{carrcasimir} we can only pose the existence of a (2,0) metric ($\eta_p^{\mu\nu}$) of the Carrollian momentum space. So we shall use this metric i.e.  $\eta_p^{\mu\nu}$ as deduced from  \eqref{carrcasimir} for further calculations. Thus, it follows that a Carrollian particle manifests a one-dimensional differentiable momentum space along the energy axis while retaining a topological dimensionality of four.  Equivalently it is only the time axis of the momentum space which inherits the deformed metric.
	\\ \\
	Now using the tetrads ($E(P_0)$) of \eqref{tetradcarroll_2}, one can find the only non-vanishing component of the momentum space metric as,
	\begin{align*}
		g_p^{00}= \eta_p^{ab} \tensor{(E)}{_a^0}
		\tensor{(E)}{_b^0}=\eta_p^{00} \tensor{(E)}{_0^0}
		\tensor{(E)}{_0^0}=\frac{1}{\left(1 + \mathfrak{a}^2 P_0^2 \right)}
	\end{align*}
	We should interpret this metric as diffeomorphism of the flat metric $\eta^{00}_p$. Now using the above one dimensional metric we can find the geodesic distance in the $p_0$ space between a chosen origin and an arbitrary point $P_0$ as
	\begin{align}	
		\nonumber
		D(0,P_0) =
		\int_{0}^{|P_0|}
		\sqrt{g_p^{00}(p_0)
			\,
			(dp_0)^2}
		=
		\int_{0}^{|P_0|} \sqrt{g_p^{00}
			(p_0)}
		\, \,
		dp_0
		&= \int_{0}^{|P_0|}
		\frac{1}{\sqrt{1+\mathfrak{a}^2(p_0)^2}}
		\,\, dp_0
		\\
		\label{g10}
		&=
		\frac{1}{\mathfrak{a}}\sinh^{-1}\Big(\mathfrak{a}|P_0|\Big)
	\end{align}
	where $|P_0|$ is identified as the particle's energy $E$. Now, we can identify $C(P_0^2)=D^2$ as the deformed Casimir of the $\kappa$-Carrollian algebra which produce a consistent commutative limit and reduces to $(P_0)^2$ as $\mathfrak{a} \to 0$. Note that we can interpret $C(P_0^2)$ as $\Pi_0^2$, in  the spirit of   \cite{biswajit2023symmetries}, where $\Pi_0$ is the zeroth component of the momentum representing the Riemann  normal coordinates defined at the tangent plane $T_O(\mathcal{P})$ anchored at the origin $O$ of $\mathcal{P}$, where this $\mathcal{P}$ now denotes 1D energy space. On the other hand the corresponding spatial normal coordinates $\vec{\Pi}(\mathfrak{a})$ can be taken as suitable deformation of $\vec{P}$ so that the dispersion relation in the normal coordinates can be recast in the limit $c\to 0$ as
	\begin{equation}
		\Pi_0 \to \left|\vec{\Pi}\right| c 
	\end{equation}				
	where $|\vec{\Pi}| \,\to |\vec{P}| $ as $\mathfrak{a}\,\to 0$ and we get the usual Carrollian dispersion relation \eqref{dispersion carroll} in the commutative limit. It is noteworthy that the functional form of the deformed d'Alembertian operator remains formally the same for both the non-relativistic and ultra-relativistic limits of $\kappa$-Minkowski space-time, except for their arguments: in the non-relativistic scenario, the operator depends on spatial momentum, whereas in the ultra-relativistic case, it solely depends on temporal component i.e. energy. And, from this, we can state that the Carrollian momentum space undergoes a reparameterisation i.e. a one-dimensional diffeomorphism only along the direction of the Carrollian temporal momentum ($P_0$), corresponding to the introduction of non-commutativity in the Carrollian space-time $\left(\widehat{\mathcal{C}}\right)$. Again, this non-linear transformation is the indication of deformed addition rule of energy. But this cannot, of course, generate curvature in 1D space; it remains flat.				
	\section{Conclusion}
	In this paper we have first discussed the non-relativistic and ultra-relativistic limits of Poincare algebra which is the isometry of flat Minkowski space time, giving rise to Galilean and Carrollian algebra. Our approach introduces a novel methodology, where space-time coordinate scaling is systematically applied, followed by the imposition of 
	$c\to\infty$ and $c\to 0$ limits. This intricate process serves to define Galilean and Carrollian coordinates, offering a unique perspective on these coordinate systems. Employing the same methodology, we extend our analysis to derive the representation of the symmetry algebra generators corresponding to each case and shown them to satisfy the correct Galilean or Bargmann and Carrollian algebra. Expanding the scope of our investigation, we redirect our focus towards exploring the ramifications of the non-relativistic and ultra-relativistic limits within the context of $\kappa$-deformed space-time (non-commutative space-time satisfying Lie algebric non-commutativity). This exploration unfolds a promising avenue to examine the interplay between the non-commutative parameter $a^\mu$, called collectively simply as $\kappa$ and the speed of light $c$ within the framework of a quantum gravity theory modeled by $\kappa$-deformed space-time. Notably, an interesting observation emerges: despite the anticipated separation of the spatio-temporal sectors in both Galilean and Carrollian limits of $\kappa$-Minkowski space-time, the persistent existence of space-time non-commutativity becomes evident. This observation aligns with earlier demonstrations highlighted in   \cite{gubitosi,symmetry}, reinforcing the resilience of non-commutative properties even under specific limits within the $\kappa$-deformed space-time paradigm. The contrasting features arising in two distinct limits of $\kappa$-deformed Minkowski spacetime become more pronounced as we delve into the calculation of the non-canonical mapping (Bopp map) between non-commutative coordinates and their commutative counterparts in each scenario. We note that to get a sensible Galilean limit of the Bopp map, the non-commutative parameter $a^{\mu}$, a set of four scalar parameters become purely space-like. On the other hand, in the Carrollian limit the parameter becomes purely time-like giving rise to a distinct time-scale. We would like to stress that, unlike   \cite{gubitosi}, where the commutative limits of both $\kappa$-Galilean and $\kappa$-Carrollian brackets produce $\kappa$-non-commutativity in only the spatio-temporal sector ($[x^0,x^i]=\frac{i}{\kappa}x^i, \,[x^i,x^j]=0$,  putting the AdS parameter $\Lambda=0$), we have achieved different space-time algebra for different limits of $\kappa$-Minkowski space-time algebra. Starting from the most general form of $\kappa$-Minkowski algebra ($[x^{\mu},x^{\nu}]=a^{\mu}x^{\nu}-a^{\nu}x^{\mu}$), we end up with the algebra (\ref{coordinate1}, \ref{coordinate2}) for Galilean limit where only spatial component of the parameter $a^{\mu}$ exists and for Carrollian limit we get only spatio-temporal non-commutativity and commutative spatial sector \eqref{def_carr_m_eq6} where only temporal component of $a^{\mu}$ exists. Interestingly, these findings aligns with the observations made in \cite{tomasz} for 2+1 dimensional $\kappa$-Minkowski space-time  where the author has shown that Carroll limit leads to a decoupling between points of space and makes the \emph{space-like} deformation somehow milder, while the absolute nature of time in the Galilei limit neutralizes the \emph{time-like} deformation.\\ 
	Finally, we explore the momentum space geometry associated with the respective algebras, leveraging the deformed phase-space algebra derived through the non-canonical mapping between commutative and non-commutative coordinates. This analysis brings forth a notable contrast in the geometric features.
	In the context of the Galilean case, we observe that the metric aspects of the momentum space are discernible exclusively in the 3-dimensional spatial sector. On the other hand, in the Carrollian scenario, the metric of the momentum space manifests solely in the one temporal dimension. In either case the structure of the metric is read off from the structures of the respective Casimir operators.
	\\
	We then, computed the geodesic distances between a chosen origin and any arbitrary point within those momentum spaces. The squared geodesic distance is then interpreted in terms of the deformed Casimir of the $\kappa$-deformed Galilean and Carrollian spaces. Intriguingly, in both instances, the deformed Casimir appears to be described by the same function. The only difference lies in their functional dependency on the respective arguments. In Galilean scenario it completely depends on the spatial components of the momentum, whereas for Carrollian space the deformed Casimir is a function of only temporal component of momentum (i.e. the energy). 
	In summary, our study not only delves into the various limits of the Poincare algebra, unraveling the Galilean and Carrollian cases, but also extends to the intricate realm of $\kappa$-deformed space-time, shedding light on the interplay between non-relativistic and ultra-relativistic dynamics in this non-commutative space-time scenario. Next we discuss some further implications and future prospects of our work. \\ \\
	Non-commutative coordinates appear quite naturally in Quantum Hall effect (QHE) in condensed matter physics \cite{Pasquier:2007nda}. Specifically, spatial noncommutativity, with time regarded as an ordinary commutative parameter, has found application in various condensed matter phenomena like anomalous quantum Hall effect, spin Hall effect \cite{horvathy2006non,haldane2018origin,Sinova:2004zz}. Interestingly, it was demonstrated in \cite{Lukierski:1996br,Duval:2001hu,Duval:2000xr} that an ``exotic" particle model, characterized by non-commuting position coordinates and linked to the two-parameter central extension of the planar Galilei group, can be employed to derive the ground states of the Fractional Quantum Hall Effect. Similarly, recent findings in \cite{hall_motions_carr_dynamics} revealed that in 2 spatial dimensions, the Carroll algebra also admits two central extensions, one of which is proportional to the non-commutative parameter of a certain type of emergent non-commutative space which are \emph{not} of the Lie-algebraic type, unlike the $\kappa$-Minkowski spacetime we are considering here \eqref{def_minkow_eq1}. Interestingly, it has been shown in \cite{hall_motions_carr_dynamics} that this kind of non-commutativity can be responsible for generating the Anomalous Hall effect even in the absence of a magnetic field. Naturally, it will be of considerable interest to investigate whether one can obtain some  form of emergent   $\kappa$-Galilean and/or $\kappa$-Carrollian particles, but now living in $(2+1)$ dimensions, with its natural Lie-algebraic type of non-commutativity and to see whether they too admit such double central extensions. This should pave the way for studying its potential applications in condensed matter system like in Hall effect, as mentioned previously. \\
	Finally, we would like to mention that in the literature   \cite{inonu1953contraction,Weimar-deformation,deAzcarraga:2002xi,Andrianopoli:2013ooa,Lukierski:2010dy} and recently in   \cite{Gomis:2019nih}, the generalized In\"{o}nu-Wigner contractions has been carried out for Lie algebras by employing direct product with several abelian semi-groups resulting in new Lie algebras (which includes graded Lie-algebras i.e. super-algebras) for different choices of the semi-group, having potentials to reveal new physical insights. In most cases the symmetry algebras describing the non-commutative spacetime are generically of Hopf algebraic type. However, in our case, apart from the commutators among the non-Lorentzian generators(by this we mean Galilean and Carrollian generators) themselves, the commutators between any of the non-Lorentzian generators and the non-commutative space-time coordinates gets deformed satisfying \eqref{phasespaceg}, \eqref{eq5}, \eqref{deformedcarroll}, \eqref{eq13}. Since the undeformed part consisting of the Galilean \eqref{mgal}, \eqref{galalg} and Carrollian algebras \eqref{rev_carr_alg_eq2} could be  obtained alternatively by the semi-group expansion method   as in \cite{Gomis:2019nih}, it will be interesting to see whether the same technique can be adopted for the extended algebras where $\kappa$-Galilean/$\kappa$-Carrollian algebras are augmented by the operator valued space-time coordinates fulfilling \eqref{coordinate1}, \eqref{coordinate2} and \eqref{def_carr_alg_eq6} and the mixed ones \eqref{phasespaceg}, \eqref{eq5}, \eqref{deformedcarroll}, \eqref{eq13}, as cited above. Here however, one has to confront the new emerging feature of the `branching' of the non-commutative parameters, as we have observed earlier, it will be quite non-trivial, if not completely impossible, to retrieve the original non-commutative parameters occurring in the completely relativistic case \eqref{def_minkow_eq1} by adopting this method. Further to the best of our knowledge, no attempts have been made in the literature to expand universal enveloping valued type of algebras using this semi-group approach. Therefore, it would be interesting from both a mathematical and physical perspective to formulate the expansion procedure for universal enveloping algebras involved in these kind of non-commutative space-times.\\ \\
	Last but not the least, we would like to mention that one can try to extend this work to the level of multi-particle systems in the $\kappa$-Galilean and $\kappa$-Carrollian background to see the effect of deformed phase space algebra in the momentum addition laws in context of particle collisions and look for some robust features of relative locality \cite{smolin} in both these limiting cases.
	\section*{Acknowledgments}
	
	D.B. extends heartfelt appreciation to Dr. Bobby Ezhuthachan (Ramakrishna Mission Vivekananda Educational \& Research Institute) for giving an introduction to Carrollian physics. Additionally, sincere acknowledgment is extended to Dr. Axel Kleinschmidt(Max Planck Institute for Gravitational Physics in Potsdam, Germany) for offering valuable insights during the progress of the paper. A.C. acknowledges support by the grant $\#$62312 from the John Templeton Foundation, as part of the ‘The Quantum Information Structure of Spacetime’ Project (QISS). The opinions expressed in this publication are those of the authors and do not necessarily reflect the views of the respective funding organization. The authors would like to thank the referee who provided useful and detailed comments on the previous version of the manuscript.
	%References
	%	\addcontentsline{toc}{section}{References} 
	\bibliographystyle{JHEP.bst}
	\bibliography{main.bib}
	
\end{document}